\documentclass[aps, twocolumn, letterpaper, superscriptaddress]{revtex4}

\usepackage{amsmath}
\usepackage{amssymb}
\usepackage{mathrsfs}
\usepackage{xspace}		
\usepackage{graphicx}				
\usepackage{braket}					
\usepackage[utf8]{inputenc}
\usepackage{enumitem}			

\setcitestyle{numbers,square}
\usepackage{hyperref}
\hypersetup{
  colorlinks   	= true, 	
  urlcolor     	= blue, 	
  linkcolor    	= blue, 	
  citecolor   	= blue 	
}
\usepackage{ifpdf}
\ifpdf
\fi

\newcommand{\eq}[1]{(\ref{#1})}
\newcommand{\Eq}[1]{Eq.~(\ref{#1})}
\newcommand{\Eqs}[1]{Eqs.~(\ref{#1})}
\newcommand{\Fig}[1]{Fig.~\ref{#1}}
\newcommand{\Sec}[1]{Sec.~\ref{#1}}
\newcommand{\Ref}[1]{Ref.~\cite{#1}}
\newcommand{\Refs}[1]{Refs.~\cite{#1}}
\newcommand{\App}[1]{Appendix~\ref{#1}}

\newcommand{\eg}{{e.g.,\/}\xspace}
\newcommand{\ie}{{i.e.,\/}\xspace}

\newcommand{\pd}{\partial}
\newcommand{\del}{\vec{\nabla}}
\newcommand{\op}[1]{\smash{\hat{#1}}}
\newcommand{\mc}[1]{\mathcal{#1}}

\newcommand{\mcu}[1]{\mathscr{#1}}
\renewcommand{\vec}[1]{{\boldsymbol{\rm #1}}}
\newcommand{\eavr}[1]{ \boldsymbol{\langle} \!\langle #1 \rangle\! \boldsymbol{\rangle} }
\newcommand{\moysin}[1]{ \boldsymbol{\{ } \! \{#1\}\! \boldsymbol{\} } }
\newcommand{\moycos}[1]{ \boldsymbol{[} [#1] \boldsymbol{]} } 
\newcommand*\xbar[1]{%
 	\, \hbox{%
 		\vbox{%
 		\hrule height 0.3pt 
 		\kern0.5ex
 		\hbox{%
 			\kern-0.2em
 			\ensuremath{#1}%
 			\kern-0.2em
		}}}\, }

\newcommand*\xbarlong[1]{%
 	\, \hbox{%
 		\vbox{%
 		\hrule height 0.3pt 
 		\kern0.5ex
 		\hbox{%
 			\kern-0.1em
 			\ensuremath{#1}%
 			\kern-0.1em
		}}}\, }

\begin{document}


\title{Zonal-flow dynamics from a phase-space perspective}

\begin{abstract}

The wave kinetic equation (WKE) describing drift-wave (DW) turbulence is widely used in studies of zonal flows (ZFs) emerging from DW turbulence. However, this formulation neglects the exchange of enstrophy between DWs and ZFs and also ignores effects beyond the geometrical-optics limit. We derive a modified theory that takes both of these effects into account, while still treating DW quanta (``driftons”) as particles in phase space. The drifton dynamics is described by an equation of the Wigner--Moyal type, which is commonly known in the phase-space formulation of quantum mechanics. In the geometrical-optics limit, this formulation features additional terms missing in the traditional WKE that ensure exact conservation of the \textit{total} enstrophy of the system, in addition to the total energy, which is the only conserved invariant in previous theories based on the WKE. Numerical simulations are presented to illustrate the importance of these additional terms. The proposed formulation can be considered as a phase-space representation of the second-order cumulant expansion, or CE2.

\end{abstract}

\author{D.~E. Ruiz}
\affiliation{Department of Astrophysical Sciences, Princeton University, Princeton, New Jersey 08544, USA}
\author{J.~B. Parker}
\affiliation{Lawrence Livermore National Laboratory, Livermore, California 94550, USA}
\author{E.~L. Shi}
\affiliation{Department of Astrophysical Sciences, Princeton University, Princeton, New Jersey 08544, USA}
\author{I.~Y.~Dodin}
\affiliation{Department of Astrophysical Sciences, Princeton University, Princeton, New Jersey 08544, USA}
\affiliation{Princeton Plasma Physics Laboratory, Princeton, New Jersey 08543, USA}
 
 \date{\today}
 
\maketitle

\section{Introduction}
\label{sec:intro}

The formation of zonal flows (ZFs) is a problem of fundamental interest in many contexts, including physics of planetary atmospheres, astrophysics, and fusion science \cite{Gurcan:2015jy, Vasavada:2005gs, Johansen:2009jf, Kunz:2013jp, Diamond:2005br, Fujisawa:2009jc, EUROfusionConsortium:2016bk}. In particular, the interaction of ZFs and drift-wave (DW) turbulence in laboratory plasmas significantly affects the transport of energy, momentum, and particles, so understanding it is critical to improving plasma confinement. But modeling the underlying physics remains a difficult problem. The workhorse approach to describing the DW-ZF coupling is the wave kinetic equation (WKE) \cite{Diamond:2005br, Trines:2005in}, but it is limited to the ray approximation \cite{foot:ray} and, in fact, is oversimplified even as a geometrical-optics (GO) model \cite{Tracy:2014to}. That leads to missing essential physics, as was recently pointed out in \Ref{foot:Parker} and will be elaborated below. These issues can be fixed by using the more accurate quasilinear approach known as the second-order cumulant expansion, or CE2 \cite{Farrell:2003dm, Farrell:2007fq, Marston:2008gx, Srinivasan:2012im, AitChaalal:2016jx}, whose applications to DW-ZF physics were pursued in \Refs{Farrell:2009ke,Parker:2013hy,Parker:2014fc, Parker:2014tb}. However, the CE2 is less intuitive than the WKE, and its robustness with respect to further approximations remains obscure. Having an approach as accurate as the CE2 and as intuitive as the WKE would be more advantageous.

Here, we propose such an approach for a DW turbulence model based on the generalized Hasegawa--Mima equation (gHME) \cite{Krommes:2000ec,Smolyakov:1999jk}. The idea is as follows. We start by splitting the gHME into two coupled equations that describe ZFs and fluctuations, respectively, and then linearize the equation for fluctuations, like in the CE2 approach. We notice then that this linearized equation is similar to that describing a quantum particle that is governed by a generalized (non-Hermitian) Hamiltonian. By drawing on this analogy, we then formulate an \textit{exact} (modulo quasilinear approximation) kinetic equation for such particle, which is akin to the so-called Wigner--Moyal equation in quantum mechanics \cite{Moyal:1949gj, Groenewold:1946kp,foot:Mendonca}. 

Compared to the CE2, the Wigner--Moyal formulation is arguably more intuitive, namely, for two reasons: (i)~like the traditional WKE (hereafter denoted by tWKE), it permits viewing DW quanta (``driftons'') as particles, except now driftons are \textit{quantumlike} particles, \ie have nonzero wavelengths; and (ii) the separation between Hamiltonian effects and dissipation remains transparent and unambiguous even beyond the GO approximation. Compared to the tWKE, the new approach is also more accurate because (i)~it captures effects beyond the GO limit, and (ii)~\textit{even in the GO limit}, it predicts corrections to the tWKE that emerge from the newly found corrections to the drifton dispersion. (In this aspect, our paper can be understood as an expansion of the GO approximation introduced in \Ref{foot:Parker}.) These corrections are essential as they allow DW-ZF enstrophy exchange, which is not included in the tWKE. By deriving the GO limit from first principles, we eliminate this discrepancy and obtain a formulation that exactly conserves the total enstrophy (as opposed to the DW enstrophy conservation predicted by the tWKE) and the total energy, in precise agreement with the underlying gHME. We also illustrate the substantial difference between the GO limit of our formulation and the tWKE using numerical simulations.

The paper is organized as follows. In \Sec{sec:basic} we introduce the gHME and its quasilinear approximation. In \Sec{sec:Wigner_Moyal} we derive the Wigner--Moyal formulation. In \Sec{sec:growth} we rederive the dispersion relation for the linear growth rate of ZFs. In \Sec{sec:WKE} we derive a corrected WKE that, in contrast to the tWKE, conserves both the total enstrophy and energy. Numerical simulations are presented to compare the new WKE with the tWKE. In \Sec{sec:conclusions} we summarize our main results. Auxiliary calculations are presented in Appendices. This includes a brief introduction to the Weyl calculus that we extensively use in our paper (\App{app:Weyl}), a spectral representation of our formulation (\App{app:spectral}), and proofs of the conservation properties of our models (\App{app:cons}).

\section{Basic model}
\label{sec:basic}

Our formulation is based on the gHME \cite{Krommes:2000ec,Smolyakov:1999jk},
\begin{gather}\label{eq:hm}
	\pd_t w + \vec{v}\cdot \del w + \beta\, \pd_x \psi = Q,
\end{gather}
which is widely used to describe electrostatic two-dimensional (2-D) turbulent flows both in a magnetized plasma with a density gradient and in an atmospheric fluid on a rotating planet, where the role of DWs is played by Rossby waves \cite{Gurcan:2015jy, Parker:2014tb}. Both contexts will be described on the same footing, so our results are applicable to DWs and Rossby waves equally. We assume the usual geophysical coordinate system, where $\vec{x} = (x, y)$ is a 2-D coordinate, the $x$-axis is the ZF direction, and the $y$-axis is the direction of the local gradient of the plasma density or of the Coriolis parameter. (In the context of fusion plasmas, a different choice of coordinates is usually preferred in literature, where $x$ and $y$ are swapped.) The constant $\beta$ is a measure of this gradient. The function $\psi(\vec{x}, t)$ is the electric potential or the stream function, $\vec{v} = \vec{e}_z \times \del \psi$ is the fluid velocity on the $\vec{x}$ plane, and $\vec{e}_z$ is a unit vector normal to this plane. The function $w(\vec{x}, t)$ is the generalized vorticity given by $\smash{w \doteq (\nabla^2 - L_{\rm D}^{-2} \op{\alpha})\psi}$, where $\op{\alpha}$ is an operator such that $\op{\alpha} = 1$ in parts of the spectrum corresponding to DWs and $\op{\alpha} = 0$ in those corresponding to ZFs. (The symbol $\doteq$ denotes definitions.) Also, $L_{\rm D}$ is the plasma sound radius or the deformation radius. (For plasmas, one can take $L_{\rm D} = 1$ in normalized units \cite{Krommes:2000ec}. Also, the barotropic model used in geophysics is recovered in the limit $L_{\rm D} \to \infty$ \cite{Farrell:2003dm,Farrell:2007fq,Marston:2008gx,Srinivasan:2012im}.) The term $Q(\vec{x}, t)$ describes external forces and dissipation. Systems with $Q = 0$ will be called isolated.

Let us decompose the fields into their zonal-averaged and fluctuating components, denoted with bars and tildes, respectively. (For any $g$, its zonal average is $\smash{\bar{g} \doteq \int \mathrm{d}x \, g /L_x}$, where $L_x$, henceforth assumed equal to one, is the system length along $x$ axis.) In particular, $w = \bar{w} + \widetilde{w}$, where the two components of the generalized vorticity are related to $\psi$ as \cite{Parker:2013hy}
\begin{gather}
	\bar{w} = \nabla^2 \bar{\psi}, 
	\quad 
	\widetilde{w} = \nabla_{\rm D}^2 \widetilde{\psi},
\end{gather}
and $\nabla_{\rm D}^2 \doteq \nabla^2 - L_{\rm D}^{-2}$. Equations for $\widetilde{w}$ and $\bar{w}$ are obtained by taking the zonal-average and fluctuating parts of \Eq{eq:hm}. This gives
\begin{subequations} \label{eq:QL1}
	\begin{gather}
		\pd_t \widetilde{w} 
			+ \widetilde{\vec{v}} \cdot \del \bar{w} 
			+ \bar{\vec{v}} \cdot \del \widetilde{w} 
			+ \beta\,\pd_x \widetilde{\psi} 
			+ f_{\rm NL}= \widetilde{Q},			
			\label{eq:w1}\\
		\pd_t \bar{w} 
			+ \overline{\widetilde{\vec{v}} \cdot \del \widetilde{w}} = \bar{Q},
			\label{eq:q1}
	\end{gather}
\end{subequations}
where $\smash{f_{\rm NL} \doteq \widetilde{\vec{v}} \cdot \del \widetilde{w} - \overline{\widetilde{\vec{v}} \cdot \del \widetilde{w}}}$ is a term nonlinear with respect to fluctuations. As discussed in \Ref{Srinivasan:2012im}, this term represents ``eddy-eddy" interactions and is responsible for the Batchelor--Kraichnan inverse-energy cascade; however, it is inessential for the formation of ZFs. Since the main scope of this paper is to specifically study the interaction between eddies and ZFs, we ignore eddy-eddy interactions so $f_{\rm NL}$ will be neglected. Hence,
\begin{subequations}	\label{eq:QL2}
	\begin{gather}
		\pd_t \widetilde{w} 
			+ \widetilde{\vec{v}} \cdot \del \bar{w} 
			+ \bar{\vec{v}} \cdot \del \widetilde{w} 
			+ \beta\,\pd_x \widetilde{\psi} 
			= \widetilde{Q},		\label{eq:w2}\\
		\pd_t \bar{w} 
			+ \overline{\widetilde{\vec{v}} \cdot \del \widetilde{w}} 
			= \bar{Q}.				\label{eq:q2}
	\end{gather}
\end{subequations}
Equations \eq{eq:QL2} compose the well-known quasilinear model \cite{Farrell:2003dm}.
In isolated systems, both sets of equations conserve the enstrophy $\mc{Z}$ and the energy $\mc{E}$ (strictly speaking, free energy), which are defined as
\begin{gather}\label{eq:ZE}
	\mc{Z} \doteq \frac{1}{2}\int \mathrm{d}^2x \,	w^2	 ,
	\quad 
	\mc{E} \doteq - \frac{1}{2} \int \mathrm{d}^2x \, w \psi .
\end{gather}

It is convenient to rewrite \Eqs{eq:QL2} in terms of the ZF velocity $\bar{\vec{v}} = \vec{e}_x U$, whose only component $U(y, t)$ is $U = - \pd_y \bar{\psi}$. Specifically, one has $\smash{\widetilde{\vec{v}} \cdot \del \bar{w} = -(\pd_x \widetilde{\psi})(\pd^2_y U)}$, $\smash{\bar{\vec{v}} \cdot \del \widetilde{w}} = U \pd_x \widetilde{w}$, and $\smash{\overline{\widetilde{\vec{v}} \cdot \del \widetilde{w}} = - \pd_y^2\, \overline{\widetilde{v}_x \widetilde{v}_y}}$. We will also assume $\widetilde{Q} = \widetilde{\xi} - \mu_{\rm dw} \widetilde{w}$ and $\smash{\bar{Q} = - \mu_{\rm zf} \bar{w}}$. Here, $\widetilde{\xi}$ is some external force with zero zonal average (eventually, we will assume it to be a white noise), and the constant coefficients $\mu_{\rm dw}$ and $\mu_{\rm zf}$ are intended to emulate the dissipation of DWs and ZFs caused by the external environment. Then, \Eqs{eq:QL2} become
\begin{subequations}	\label{eq:QL3}
	\begin{gather}
		\pd_t \widetilde{w} 
				+ U \pd_x \widetilde{w} 
				+ [\beta - (\pd_y^2 U)] \pd_x \widetilde{\psi} 
				= \widetilde{\xi} - \mu_{\rm dw} \widetilde{w},
				\label{eq:w3}\\
		\pd_t U 
				+ \mu_{\rm zf} U 
				+ \pd_y \overline{\widetilde{v}_x \widetilde{v}_y} = 0.
				\label{eq:q3}
	\end{gather}
\end{subequations}

Equations \eq{eq:QL3} are the same model as the one that underlies the CE2. Although not exact, this model is useful because it captures key aspects of ZF dynamics, such as formation and merging of zonal jets \cite{Srinivasan:2012im, Parker:2014tb, Constantinou:2014fh}. Below, we use it to derive a formulation of DW-ZF interactions alternative to the CE2.

\section{Wigner--Moyal formulation}
\label{sec:Wigner_Moyal}

\subsection{State vector}

Consider a family of all reversible linear transformations of $\widetilde{w}(\vec{x}, t)$ of the form $\smash{\int \mathrm{d}^2x' \,	K(\vec{x}, \vec{x}', t) \widetilde{w}(\vec{x}', t)}$. These transformations map $\widetilde{w}(\vec{x}, t)$ into some family of image functions. Since these functions are mutually equivalent up to an isomorphism, the resulting family can be viewed as a single object, a time-dependent ``state vector'' $\ket{\widetilde{w}}$. (Analogous definitions will be assumed also for $\smash{\ket{\widetilde{\psi}}}$ and $\smash{\ket{\widetilde{\xi}}}$.) The original function $\widetilde{w}(\vec{x},t)$ is then understood as a projection of $\ket{\widetilde{w}}$, namely, as its ``coordinate representation'' given by $\widetilde{w}(\vec{x}, t) = \braket{\vec{x} | \widetilde{w}}$. Here, $\ket{\vec{x}}$ are the eigenstates of the position operator $\op{\vec{x}}$ normalized such that $\braket{ \vec{x}' |\op{\vec{x}} | \vec{x}}= \vec{x} \braket{\vec{x}'|\vec{x}}= \vec{x}\, \delta(\vec{x}' - \vec{x})$. This definition of a field is similar to that used in quantum mechanics for describing probability amplitudes \cite{foot:qm}. Hence, it is convenient to describe the dynamics of $\ket{\widetilde{w}}$ using a quantumlike formalism. This is done as follows.

In addition to the coordinate operator $\op{\vec{x}}$, we introduce a momentum (wave-vector) operator $\op{\vec{p}}$ such that, in the coordinate representation, $\op{\vec{p}} \doteq - i\del$. Accordingly, $\ket{\widetilde{w}} = - \op{p}_{\rm D}^2 \ket{\widetilde{\psi}}$, where
\begin{gather}
 	\op{p}_{\rm D}^2 \doteq \op{p}^2 + L_{\rm D}^{-2}, 
 	\quad 
 	\op{p}^2 \doteq \op{\vec{p}} \cdot \op{\vec{p}}.
 	\label{eq:pbar}
\end{gather}
Hence, \Eq{eq:w3} can be represented in the following form:
\begin{gather}
 	i \pd_t \ket{\widetilde{w}} 
  	= \op{H} \ket{\widetilde{w}} + i \ket{\widetilde{\xi}}.
 \label{eq:omega}
\end{gather}
The operator $\op{H}$ is given by
\begin{gather}\label{eq:H}
	\op{H} \doteq 
		- \beta \op{p}_x \op{p}_{\rm D}^{-2} 
     	+ \op{U} \op{p}_x 
      	+ \op{U}'' \op{p}_x \op{p}_{\rm D}^{-2}
      	- i \mu_{\rm dw}.
\end{gather}
Also, $\smash{\op{U} \doteq U(\op{y},t)}$, and the prime above $U$ henceforth denotes $\pd_y$; in particular, $\smash{\op{U}'' \doteq \pd_y^2 \, U(\op{y},t)}$.

\subsection{Generalized von Neumann equation}

Let us express \Eq{eq:H} as $\smash{\op{H} = \op{H}_H + i \op{H}_A}$, where $\smash{\op{H}_H \doteq (\op{H} + \op{H}^\dag)/2}$ and $\smash{\op{H}_A \doteq (\op{H} - \op{H}^\dag)/(2i)}$ are the Hermitian and anti-Hermitian parts of $\op{H}$, correspondingly. Explicitly, these operators can be written as
\begin{subequations}
 	\begin{gather}
 		\op{H}_H 
 			= - \beta \op{p}_x \op{p}_{\rm D}^{-2} 
       		+ \op{U} \op{p}_x 
       		+ [\op{U}'' , \op{p}_x \op{p}_{\rm D}^{-2} ]_+/2,    
       			\label{eq:D_H}  \\
  		\op{H}_A 
  			= [\op{U}'' , \op{p}_x \op{p}_{\rm D}^{-2} ]_- / (2i) 
  			- \mu_{\rm dw},  
  			\label{eq:D_A}
	\end{gather}
\end{subequations}
where $[\cdot, \cdot ]_-$ denotes the commutator given by $\smash{[\hat{A},\hat{B}]_-= \hat{A} \hat{B} - \hat{B}\hat{A}}$ and $[\cdot, \cdot ]_+$ denotes the anti-commutator given by $\smash{[\hat{A},\hat{B}]_+= \hat{A} \hat{B} + \hat{B}\hat{A}}$. Let us also introduce a Hermitian operator $\smash{\op{W} \doteq \ket{\widetilde{w}}\bra{\widetilde{w}}}$, which, by analogy with quantum mechanics, is interpreted as the ``fluctuating-vorticity density'' operator. It is seen from \Eq{eq:omega} that $\op{W}$ satisfies
\begin{gather}
 	i\pd_t \op{W} 
 		= [\op{H}_H, \op{W}]_- 
 		+ i[\op{H}_A, \op{W}]_+
 		+ i\op{F},
 \label{eq:Neumann}
\end{gather}
where $\smash{\op{F} \doteq \ket{\widetilde{\xi}} \bra{\widetilde{w}} + \ket{\widetilde{w}} \bra{\widetilde{\xi}}}$. In particular, taking the trace of this equation also gives an equation for the ``total number of DW quanta,'' $N \doteq \text{Tr}\,\op{W} = \int \mathrm{d}^2 x \, \braket{ \vec{x} | \hat{W} | \vec{x} }   = \int \mathrm{d}^2 x \, \widetilde{w}^2 = \braket{\widetilde{w} | \widetilde{w}} $; namely,
\begin{gather}
	\dot{N} = 2\text{Tr}\,(\op{H}_A \op{W}) + \text{Tr}\,\op{F}.
\end{gather}
This indicates that $\op{H}_A$ determines the loss of quanta, or dissipation of DWs. [In particular, the term $\mu_{\rm dw}$ in \Eq{eq:D_A} is responsible for DW dissipation to the external environment, whereas the term $[\op{U}'' , \op{p}_x \op{p}_{\rm D}^{-2} ]_- / (2i) $ destroys DW quanta while conserving the total enstrophy, as will be discussed in \Sec{sec:maineq}.] Also, $\op{H}_H$ determines conservative dynamics of DWs and thus can be understood as the \textit{drifton Hamiltonian}. (The non-Hermitian operator $\op{H}$ will be attributed as the generalized Hamiltonian.) Notice that the distinction between dissipation and Hamiltonian effects remains unambiguous even beyond the GO approximation.

Equation \eq{eq:Neumann} can be understood as a generalized von~Neumann equation akin to the one that commonly emerges in quantum mechanics. A standard approach to such equation is to project it on the phase space using the Weyl transform. Hence, we proceed as follows. (Readers who are not familiar with the Weyl calculus are encouraged to read \App{app:Weyl} before continuing further.)

\subsection{Wigner--Moyal equation}
\label{sec:wm}

Let us introduce $W$ as the Weyl symbol of $\op{W}$, \ie
\begin{gather}
 	W(\vec{x}, \vec{p}, t) 
 		\doteq \int \mathrm{d}^2 s \, 
 		e^{-i \vec{p} \cdot \vec{s}} 
    		\braket{\vec{x} + \vec{s}/2| \op{W} | \vec{x} - \vec{s}/2},
\end{gather}
which is real because $\op{W}$ is Hermitian. In quantum mechanics, a similar construct is known as the Wigner function \cite{Wigner:1932cz}, so one can readily identify the physical meaning of $W$. Specifically, in the regime when the ray approximation applies and dissipation is negligible, $W/(2\pi)^2$ represents the phase-space probability density of driftons [the numerical coefficient $(2\pi)^2$ comes from \Eq{eq:trace}], while beyond the GO limit it can be considered as a \textit{generalization} of this probability density \cite{foot:quasi}. Using the fact that $\widetilde{w}(\vec{x}, t)$ is real, one can also cast $W$ as
\begin{gather}\label{eq:auxW}
 	W(\vec{x}, \vec{p}, t) 
 		\doteq \int \mathrm{d}^2 s \, 
 		e^{-i \vec{p} \cdot \vec{s}}\,
    		\widetilde{w}\!\left(\vec{x} + \frac{\vec{s}}{2}, t\right) 
    		\widetilde{w}\!\left(\vec{x} - \frac{\vec{s}}{2}, t\right),
\end{gather}
which also implies
\begin{gather}\label{eq:rW}
	W(\vec{x}, \vec{p}, t) = W(\vec{x}, -\vec{p}, t).
\end{gather}
One can interpret the right-hand side of \Eq{eq:auxW} as the local spatial spectrum of the correlation function of $w$. Hence, $W$ will be called the \textit{DW spectral function}.

By applying the Weyl transform to \Eq{eq:Neumann}, one obtains the following pseudo-differential equation \cite{foot:pseudo}:
\begin{gather}\label{eq:aux1}
	\pd_t W = \moysin{H_H,W} + \moycos{H_A, W} + F. 
\end{gather}
Here $\moysin{\cdot,\cdot}$ and $\moycos{\cdot, \cdot}$ are Moyal's ``sine bracket'' [\Eq{eq:sine_bracket}] and ``cosine bracket'' [\Eq{eq:cosine_bracket}]. The functions $H_H(y,\vec{p},t)$, $H_A(y,\vec{p},t)$, and $F(\vec{x},\vec{p},t)$ are the Weyl symbols of $\op{H}_H$, $\op{H}_A$, and $\op{F}$, respectively. In particular, using \Eq{eq:Moyal} and the fact that $U$ is independent of $x$, one obtains
\begin{gather}
	H_H = - \beta p_x/p_{\rm D}^2 
			+ p_x  U  + \moycos{ U'' , p_x / p_{\rm D}^2}/2, \\
	H_A = \moysin{ U'', p_x / p_{\rm D}^2} /2 
			- \mu_{\rm dw},
\end{gather}
where $p_{\rm D}^2 \doteq p^2 + L_{\rm D}^{-2}$. By analogy with quantum mechanics, we call \Eq{eq:aux1} a Wigner--Moyal equation.

Next, let us consider the zonal average of this equation, 
\begin{gather}
	 \pd_t \xbar{W} = 
    		\moysin{H_H, \xbar{W} } 
    		+ \moycos{H_A, \xbar{W} } 
    		+ \xbarlong{F},
 	\label{eq:aux2}
\end{gather}
where $\xbar{W} = \xbar{W}(y, \vec{p}, t)$. We adopt the ergodic assumption, namely, that the zonal average is equivalent to the ensemble average [denoted $\smash{\eavr{\ldots}}$] over realizations of the random force $\widetilde{\xi}$ (\eg as done in \Ref{Parker:2013hy}). To calculate $\smash{\xbarlong{F} = \eavr{F}}$, consider integrating \Eq{eq:omega} on a time interval $(t_0, t)$. The result can be written as $\smash{
		\ket{\widetilde{w}_t} =
				 \ket{\widetilde{w}_{t_0}} 
				 + \ket{\delta \widetilde{w}_t} 
				 + \int_{t_0}^t \mathrm{d} t' \ket{\widetilde{\xi}_{t'}},
}$ where the indexes denote the times at which functions are evaluated, and $\smash{\ket{\delta \widetilde{w}_t} \doteq - i \int_{t_0}^t \mathrm{d} t' \op{H} \ket{\widetilde{w}_{t'}}}$. We assume
\begin{gather}
	\eavr{ \widetilde{\xi}(\vec{x},t) \widetilde{\xi}(\vec{x}',t')} 
			= \delta(t-t')\,\,\Xi \boldsymbol{(} (y + y')/2, \vec{x}-\vec{x}' \boldsymbol{)},
\end{gather}
where $\Xi$ is a correlation function that is homogeneous in $x$ but not necessarily in $y$ \cite{Srinivasan:2012im,Parker:2014tb}. Since $\smash{\ket{\delta \widetilde{w}_t}}$ can be affected by $\ket{\widetilde{\xi}_{t'}}$ only if $t > t'$, one has $\smash{\eavr{\ket{\widetilde{\xi}_t}\bra{\delta \widetilde{w}_t}}} = 0$. Hence,
\begin{align}
\xbarlong{F}(y, \vec{p})
	=  &  \int \mathrm{d}^2 s \,
      				e^{-i \vec{p} \cdot \vec{s}} 
      				\eavr{
      					\bra{\vec{x}+\frac{\vec{s}}{2} }  
      					( \ket{ \tilde{\xi}_t} \bra{ \tilde{w}_t } + \text{h.c.} ) 
      					\ket{ \vec{x}-\frac{\vec{s}}{2} } 
      				}   \notag \\
	=  &  \int \mathrm{d}^2 s \, 
      				e^{-i \vec{p} \cdot \vec{s}} \int_{t_0}^t \mathrm{d}t' \, 
      				\delta(t-t') 
      				\left[ \Xi(y, \vec{s}) + \Xi(y, -\vec{s}) \right] 
      				  \notag \\
	=  &  \frac{1}{2} \int \mathrm{d}^2 s \, 
      				e^{-i \vec{p} \cdot \vec{s}}
      				\left[ \Xi(y, \vec{s}) + \Xi(y, -\vec{s}) \right] 
      				  \notag \\
	=  &  \int \mathrm{d}^2 s \, \,
		      \Xi(y, \vec{s}) \cos(\vec{p} \cdot \vec{s}),
\end{align}
where `$\text{h.c.}$' denotes ``Hermitian conjugate.'' In other words, once the correlation function $\Xi$ of $\widetilde{\xi}$ is specified, $\xbarlong{F}$ can be readily calculated as the Fourier image of $\Xi$. 

This concludes the calculation of the functions that determine the evolution of $\xbar{W}$ through \Eq{eq:aux2}. However, these functions generally depend on $U$, so an additional equation for $U$ is needed to make the theory self-consistent. This equation is derived as follows.

\subsection{Equation for the zonal-flow velocity}

Returning to \Eq{eq:q3}, we rewrite the nonlinear term~as
\begin{align}
	\widetilde{v}_x \widetilde{v}_y
 			& = - (\pd_y \widetilde{\psi}) (\pd_x \widetilde{\psi}) \notag\\
 			& = - \bra{ \vec{x}} \op{p}_y \ket{\widetilde{\psi}} 
   					\bra{ \widetilde{\psi}} \op{p}_x \ket{ \vec{x} } \notag \\
			& = - \braket{ \vec{x} |
       				\op{p}_y \op{p}_{\rm D}^{-2} \op{W} 
       				\op{p}_{\rm D}^{-2} \op{p}_x  	| \vec{x} } 
       			\notag \\
			& = - \int \frac{d^2 p}{(2\pi)^2}\,
					\frac{p_y}{ p_{\rm D}^2} \star W \star \frac{p_x}{ p_{\rm D}^2},
\end{align}
where we used \Eq{eq:weyl_x_rep2} in the last step. After introducing the averaged vorticity density $\xbar{W}$, \Eq{eq:q3} becomes
\begin{equation}
		\pd_t U + \mu_{\rm zf} U 
				 = 	\frac{\pd}{\pd y} \int \frac{\mathrm{d}^2 p}{(2\pi)^2}\,
						\frac{p_y}{p_{\rm D}^2} \star 
						\xbar{W} \star \frac{p_x}{ p_{\rm D}^2} .
		\label{eq:auxU}
\end{equation}
Since $\xbar{W}$ is independent of $x$ and satisfies the condition \eq{eq:rW}, \Eq{eq:auxU} can also be written as
\begin{equation}
		\pd_t U + \mu_{\rm zf} U 
				 = 	\frac{\pd}{\pd y} \int \frac{\mathrm{d}^2 p}{(2\pi)^2}\,
						\frac{1}{p_{\rm D}^2} \star 
						p_x p_y \xbar{W} \star \frac{1}{ p_{\rm D}^2} .
		\label{eq:auxU2}
\end{equation}
The combination of \Eqs{eq:aux2} and \eq{eq:auxU2} forms a closed set of equations that can be used to calculate the dynamics of $\xbar{W}$ and $U$ self-consistently.

\subsection{Main equations and conservation laws}
\label{sec:maineq}

Let us slightly change the notation and summarize the above equations in the following form:
\begin{subequations} \label{eq:phase_space}
	\begin{gather}  
 		\pd_t \xbar{W} =
    			\moysin{ \mc{H}, \xbar{W}} 
   			+ \moycos{ \Gamma, \xbar{W} } 
   			+ \xbarlong{F}
   			- 2\mu_{\rm dw} \xbar{W},         \label{eq:Moyal_Liouville}  \\
	 	\pd_t U+ \mu_{\rm zf} U 
  			= \frac{\pd}{\pd y}
   			\int \frac{\mathrm{d}^2 p}{(2\pi)^2}
   			\frac{1}{ p_{\rm D}^2} \star p_x p_y \xbar{W} 
   			\star \frac{1}{ p_{\rm D}^2} \label{eq:flow}.
	\end{gather}
\end{subequations}
As a reminder, $\xbar{W}(y, \vec{p}, t)$ is the zonal-averaged spectral (or Wigner) function that describes DW turbulence, and $U(y, t)$ is the ZF velocity. Also, $\smash{\xbarlong{F} = \xbarlong{F}(y, \vec{p})}$ is determined by the correlation function of the external noise $\smash{\widetilde{\xi}}$ (\Sec{sec:wm}). We have also introduced $\mc{H} \doteq H_H$ and $\Gamma \doteq H_A + \mu_{\rm dw}$, or, explicitly,
\begin{subequations} \label{eq:coefficients}
 	\begin{gather}
  		\mc{H}(y,\vec{p},t) =
    			-\beta p_x / p_{\rm D}^2  
    			+ p_x U 
    			+ \moycos{ U'' , p_x / p_{\rm D}^2} /2, \label{eq:Hamiltonian} \\
  		\Gamma(y, \vec{p},t) =
    			\moysin{ U'', p_x / p_{\rm D}^2}/2 .  \label{eq:Damping}
 \end{gather}
\end{subequations}
In \App{app:spectral}, we also present a spectral representation of these equations that can be used for a numerical implementation of the Wigner--Moyal formulation.

The function $\mc{H}$ can be understood as the Weyl symbol of the drifton Hamiltonian, whereas $\Gamma$ determines dissipation of DW quanta that is caused specifically by DW interaction with ZFs. This is explained as follows. Since \Eqs{eq:phase_space} are \textit{exact} within the quasilinear approximation (modulo the ergodic assumption), they inherit the same conservation laws as the original quasilinear model given by \Eqs{eq:QL3}. Specifically, for isolated systems ($\xbarlong{F} = 0$ and $\mu_{\rm dw, zf} = 0$), \Eqs{eq:phase_space} and \eq{eq:coefficients} exactly conserve the \textit{total} enstrophy and energy [\Eqs{eq:ZE}]
\begin{gather}\label{eq:invariants}
		\mc{Z} = \mc{Z}_{\rm dw} + \mc{Z}_{\rm zf},
		\quad
		\mc{E} = \mc{E}_{\rm dw} + \mc{E}_{\rm zf}
\end{gather}
rather than their DW and ZF components. (A direct proof is given in \App{app:ee3}.) For completeness, we present expressions for these components:
\begin{subequations}	\label{eq:invariants_II}
	\begin{align}  
		\mc{Z}_{\rm dw} 
			& 	\doteq \frac{1}{2} \int \mathrm{d}^2 x \, \widetilde{w}^2
 				= \frac{1}{2} \int \frac{\mathrm{d}^2 p}{(2 \pi)^2} \,\mathrm{d} y \, 
 					\xbar{W}, \label{eq:Zdw}\\
		\mc{Z}_{\rm zf} 
			& 	\doteq \frac{1}{2} \int \mathrm{d} y \, \bar{w}^2
 				= \frac{1}{2} \int \mathrm{d} y \, (U')^2, \label{eq:Zzf}\\
		\mc{E}_{\rm dw} 
			& 	\doteq - \frac{1}{2} \int \mathrm{d}^2 x \, \widetilde{w} \widetilde{\psi}
 				= \frac{1}{2} \int \frac{\mathrm{d}^2 p}{(2 \pi)^2}\,\mathrm{d} y \, 
 						\frac{\xbar{W}}{p_{\rm D}^2},\\
		\mc{E}_{\rm zf} 
			& 	\doteq - \frac{1}{2} \int \mathrm{d} y \, \bar{w} \bar{\psi} 
 				= \frac{1}{2} \int \mathrm{d} y \, U^2,
	\end{align}
\end{subequations}
where we used \Eqs{eq:trace} and \eq{eq:int_everywhere} to derive the second set of equalities. According to \Eqs{eq:Zdw} and \eq{eq:trace}, note that the DW enstrophy $\mc{Z}_{\rm dw}$ and the total number of DW quanta $\smash{N \doteq \text{Tr}\,\op{W}}$ are the same up to a constant factor.

The conservative equations \eq{eq:phase_space} and \eq{eq:coefficients}, which we attribute as the Wigner--Moyal formulation of DW-ZF interactions, constitute the main result of our work. This formulation can be understood as an alternative phase-space representation of the CE2 since it is derived from the same quasilinear model. However, the Wigner--Moyal formulation is arguably more intuitive than the CE2, namely, for two reasons: (i) Like in the tWKE, driftons are treated as particles, except now they are \textit{quantumlike} particles, \ie have nonzero wavelengths; hence, one is not constrained to the GO limit. (ii) Also, the separation between Hamiltonian effects and dissipation remains transparent and unambiguous even beyond the GO approximation. The Wigner--Moyal formulation elucidates the link between the WKE formalism and the CE2 and also helps make approximations rigorous by making them systematic. Below, these and other applications are discussed in further detail.

\section{Growth rate of zonal flows}
\label{sec:growth}

\subsection{Basic equations}

To demonstrate the convenience of the Wigner-Moyal formulation, let us apply it to rederive the rate of the linear zonostropic instability, \ie the growth rate of weak ZFs. Suppose a homogeneous equilibrium with zero ZF velocity and some DW spectral function $\mcu{W}(\vec{p})$. [As pointed out in \Sec{sec:wm}, the corresponding $\mcu{W}(\vec{p})/(2\pi)^2$ represents the phase-space probability distribution of driftons.] Consider small perturbations to this equilibrium; namely,
\begin{align}
 	U = \delta U(y,\vec{p},t), 
 		& \quad 
 		\delta U = \text{Re}\,(U_q e^{iqy + \gamma t}),\notag \\
 	\xbar{W} = \mcu{W}(\vec{p}) + \delta \xbar{W}(y,\vec{p},t),
 		& \quad 
 		\delta \xbar{W} = \text{Re}\,[\xbar{W}_q(\vec{p}) e^{iqy + \gamma t}].\notag
\end{align}
Here, the constant $q$ serves as the modulation wave number, and the constant $\gamma$ is the instability rate to be found. The linearization of \Eq{eq:Moyal_Liouville} leads to
\begin{multline} 
 	(\pd_t + 2\mu_{\rm dw}) \delta\xbar{W} 
    		+\moysin{ \beta p_x / p_{\rm D}^2 , \delta\xbar{W} \,} \\
    		= \moysin{  p_x \delta U , \mcu{W}} 
    		+  \moysin{ \moycos{ \delta U'', p_x/ p_{\rm D}^2}/2 , \mcu{W}} \\
     	+ \moycos{   \moysin{   \delta U'', p_x / p_{\rm D}^2}/2 , \mcu{W} } ,
\end{multline}
where we substituted \Eqs{eq:coefficients}. The brackets can be calculated using \Eqs{app:brackets_exp}. Hence, we obtain
\begin{multline}  
 	\left[ i (\gamma + 2 \mu_{\rm dw})
     		+ \beta p_x \left(\frac{1}{ p_{\mathrm{D},+q}^2} 
     		- \frac{1}{ p_{\mathrm{D},-q}^2} \right) 
  		\right] \xbar{W}_q \\
 		= (\mcu{W}_{+q} - \mcu{W}_{-q})
    		\left[  \frac{p_x q^2}{2} \left(\frac{1}{ p_{\mathrm{D},+q}^2} 
    		+ \frac{1}{ p_{\mathrm{D},-q}^2} \right) - p_x \right] U_q \\
  		+ \frac{ p_x q^2}{2}(\mcu{W}_{+q} + \mcu{W}_{-q}) 
     	\left(\frac{1}{p_{\mathrm{D},+q}^2} - \frac{1}{p_{\mathrm{D},-q}^2} \right) U_q,
\end{multline}
where we assume the notation $\smash{A_{\pm q} \doteq A(\vec{p} \pm \vec{e}_y q/2)}$ for any $A$. Solving for $\xbar{W}_q$ in terms of $U_q$ leads to
\begin{multline*}
 	\xbar{W}_q =  
 		\frac{ i p_x p_{\mathrm{D},+q}^2 p_{\mathrm{D},-q}^2}
   		{ (\gamma + 2 \mu_{\rm dw})p_{\mathrm{D},+q}^2 p_{\mathrm{D},-q}^2 
   					+ 2 i \beta q p_x p_y} \\
 	 	\times
  		\left[\mcu{W}_{+q}  \left(1- \frac{q^2}{ p_{\mathrm{D},+q}^2} \right)   
    			-  \mcu{W}_{-q}  \left(1 - \frac{q^2}{ p_{\mathrm{D},-q}^2} \right)  
   		\right] U_q .
\end{multline*}
Then, \Eq{eq:flow} yields
\begin{gather*} 
 	(\gamma + \mu_{\rm zf}) e^{iqy} U_q 
   		=  \frac{\pd}{\pd y} 
    			\int \frac{\mathrm{d}^2 p}{(2\pi)^2} \,
     		\frac{1}{ p_{\rm D}^2} \star p_x p _y e^{i q y} 
     		\xbar{W}_q \star \frac{1}{ p_{\rm D}^2}.		
\end{gather*}
Due to \Eq{app:moyal_exp}, this can be simplified as follows:
\begin{gather}  \label{eq:flow_perturbation}
 	(\gamma + \mu_{\rm zf}) U_q 
   		= i q  \int \frac{\mathrm{d}^2 p}{(2\pi)^2} 
   		\frac{p_x p_y}{ p_{\mathrm{D},+q}^2 p_{\mathrm{D},-q}^2}\, \xbar{W}_q.
\end{gather}
After substituting the expression for $\xbar{W}_q$, one gets
\begin{multline}  
	\gamma + \mu_{\rm zf} 
  		= \int \frac{\mathrm{d}^2 p}{(2\pi)^2}\, 
  			\frac{ q p_x^2 p_y}
    			{ (\gamma + 2 \mu_{\rm dw})p_{\mathrm{D},+q}^2 p_{\mathrm{D},-q}^2 
    				+ 2 i \beta q p_x p_y} \\
  		\times
  		\left[	\mcu{W}_{-q}
  				\left(1 - \frac{q^2}{ p_{\mathrm{D},-q}^2} \right)  
 				- \mcu{W}_{+q}\left(1- \frac{q^2}{ p_{\mathrm{D},+q}^2} \right)   
   		\right].
   	\label{eq:dispersion}
\end{multline}

As expected, this dispersion relation coincides with that obtained using the CE2 formalism \cite{Srinivasan:2012im}. Notably, the dependence of the integrand on $\mcu{W}_{\pm q}$ makes the expression similar to dispersion relations that emerge in quantum mechanics; for instance, cf. Ref.~\cite[Sec.~40]{Lifshitz:1981ui}.

\subsection{Zonal flows with nonzero group velocity}

\begin{figure*}
	\centering
	\includegraphics[scale=0.56]{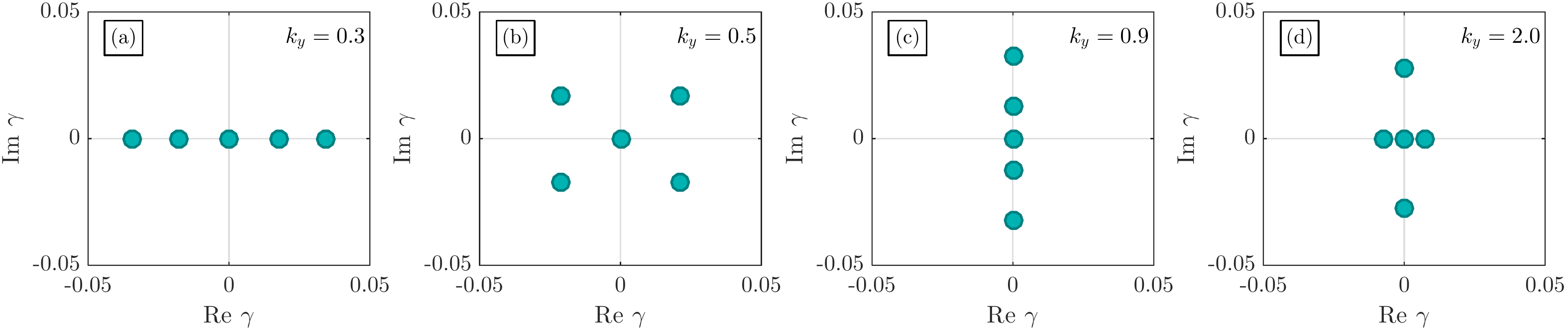}
	\caption{Numerical solutions of the dispersion relation \eq{eq:disp_rel} for different $k_y$ with fixed $k_x = 1$, $q=0.1$, $\beta=1$, $L_{\rm D}=1$, $\mu_{\rm dw, zf}=0$, and $\mcu{N}=1$. The solutions shown in subfigures (a)-(d) correspond to $k_y=0.3$, $k_y=0.5$, $k_y=0.9$, and $k_y=2.0$, respectively. In the interval $0.33  \lesssim  |k_y| \lesssim 0.82 $, the solutions $\gamma$ can be complex-valued. With the same forcing and same fixed parameters, similar regimes can also be observed in the barotropic limit $(L_{\rm D} \to \infty )$ or in the case of nonzero dissipation.
	}
	\label{fig:solutions}
\end{figure*}

As a side note, it is commonly thought that ZFs only grow in situ; \ie $\text{Re}\,\gamma > 0$ with $\text{Im}\,	\gamma = 0$. There have been questions over whether it is possible to have unstable zonal modes at nonzero $\text{Im}\,\gamma$, which implies nonzero group velocity \cite{Bakas:2015iy}. Here we show, by presenting an example, that the answer is yes. Let us consider the following steady state:
\begin{multline}
	\mcu{W} = (2\pi)^2 \mcu{N} 
					[ \delta( p_x- k_x)\delta( p_y- k_y) 
						+  \delta( p_x+ k_x)\delta( p_y + k_y) \\
						+  \delta( p_x+ k_x)\delta( p_y- k_y) 
						+  \delta( p_x- k_x)\delta( p_y+ k_y) ]/4.
\end{multline}
After integrating, \Eq{eq:dispersion} can be cast as follows:
\begin{multline}
	0 =\gamma +\mu_{\rm zf} -
  		   	\frac{ X \mcu{N} k_x^2 }{v_{gy} k_{\mathrm{D}}^4 } 
  			\left(1- \frac{q^2}{ k_{\mathrm{D}}^2 } \right) \\
  			\times	
  			\sum_{ n = -1,1 } 
  					\frac{  n( k_y+nq/2 ) 
  							 k_{\mathrm{D},+2nq}^2 / k_{\mathrm{D}}^2 }
    						{ X^2 k_{\mathrm{D},+2nq}^4 /  k_{\mathrm{D}}^4
    							+ (k_y+nq/2)^2/k_y^2 },
    	\label{eq:disp_rel}
\end{multline}
where $v_{gy} \doteq 2 \beta k_x k_y / k^4_{\rm D }$ is the DW group velocity in the absence of ZFs, $k_{\mathrm{D},\pm q}^2 \doteq k_x^2 + (k_y \pm q/2)^2 + L_{\rm D}^{-2}$, and $X \doteq (\gamma+2 \mu_{\rm dw}) / (q v_{gy})$. Numerical solutions of \Eq{eq:disp_rel} are presented in \Fig{fig:solutions}. As shown, solutions for $\gamma$ are complex over some interval of $k_y$. This counterexample shows the existence of ``traveling" unstable ZF modes.

Additional insights on this phenomenon can be obtained by considering the limit, where $\mu_{\rm dw, zf} \ll |\gamma|$ and $q \ll k_y$. In this case, \Eq{eq:disp_rel} simplifies to
\begin{equation}
	0 = X \left[ 1-8\sigma  \frac{X^2-1}{(X^2+1)^2}		\right]   + \mc{O}(q^2),
    \label{eq:disp_X}
\end{equation}
where 
\begin{equation}
	\sigma \doteq 
				\frac{\mcu{N} k_x^2 }{8 v^2_{gy} k_{\mathrm{D}}^4 } 
  				\left(1- \frac{4 k_y^2}{ k_{\mathrm{D}}^2 } \right).
  	\label{eq:disp_GO}
\end{equation}
One may consider this as the GO limit of \Eq{eq:disp_rel}. Equation \eq{eq:disp_X} predicts four nontrivial solutions for $X$, which are given by $X^2 = -1+4 \sigma \pm 4 [\sigma(\sigma-1)]^{1/2}$. Different regimes for the solutions can be deduced. When $\sigma \geq 1$, the solutions $\gamma$ are purely real. For the parameters in \Fig{fig:solutions}, this regime corresponds to $ |k_y| \lesssim 0.33 $. In the interval $0 < \sigma < 1$ corresponding to $0.33  \lesssim  |k_y| \lesssim 0.82 $, $\gamma$ is complex-valued. In the interval $-1/8 \leq \sigma \leq 0$, which corresponds to $0.82 \lesssim |k_y| \lesssim 1.07$, the solutions are purely imaginary. Finally, in the interval $\sigma < -1/8$ corresponding to $ |k_y| \gtrsim 1.07 $, two solutions $\gamma$ are purely imaginary, and two other solutions are purely real. The different regimes identified by solving \Eq{eq:disp_X} are consistent with the observed numerical solutions of the exact dispersion relation \eq{eq:disp_rel}. In the next section, we will explore the GO limit of the DW-ZF interactions in more detail.

\section{Geometrical-optics limit and the wave kinetic equation}
\label{sec:WKE}
 
 \begin{figure*}
\centering
\includegraphics[scale=0.7]{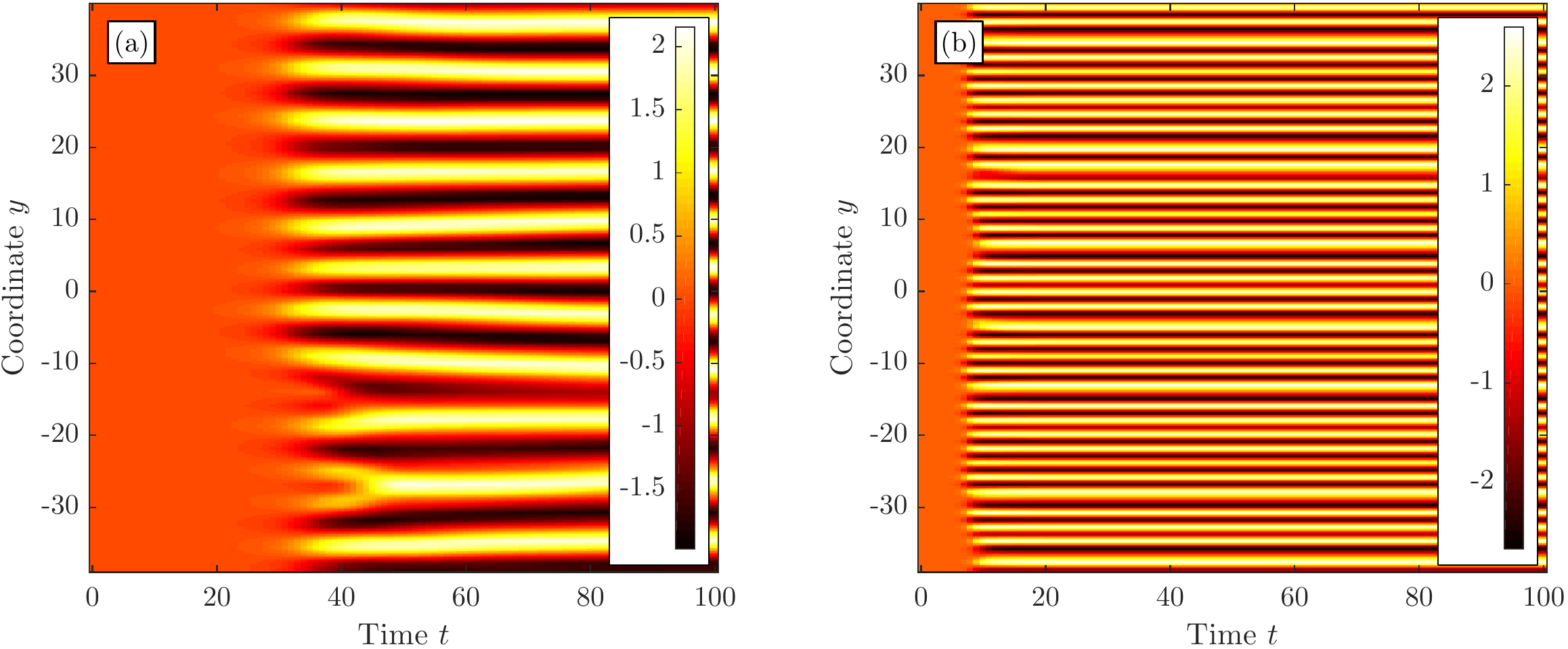}
\caption{The ZF velocity $U(y, t)$ obtained by numerically integrating the WKE \eq{eq:WKE} for $\mc{H}$ and $\Gamma$ of two types: (a) our model [\Eqs{eq:WKE_II}]; (b) the tWKE model [\Eqs{eq:WKE_III}]. Both simulations used the same parameters and initial conditions. Small initial values for $\xbar{W}$ and $U$ were randomly assigned such that \Eq{eq:rW} was satisfied. The parameters used are: $\beta = 1$, $ L_{\rm D} = 1$, $\mu_{\rm dw,zf}  = 0.1$, and $\smash{\xbarlong{F} = 4 \pi\delta(|\vec{p}|-1)}$. Equation \eq{eq:WKE_drift} was discretized in a $[-39,39] \times [-2,2] \times [-4,4]$ phase space using a discontinuous-Galerkin (DG) method \cite{Liu:2000ee} on a uniformly-spaced Cartesian grid with $80 \times 24 \times 48$ cells while \Eq{eq:WKE_zonal} was discretized on a subset of this grid. Time advancement was done using an explicit third-order strong-stability-preserving Runge-Kutta algorithm \cite{Gottlieb:2001iy}. The solution was expanded locally in each cell as a sum of piecewise polynomials of degree one. At cell interfaces, an upwind numerical flux was used in \Eq{eq:WKE_drift} and a centered numerical flux was used in \Eq{eq:WKE_zonal}. Higher-order spatial derivatives such as $U''$ and $U'''$ were computed using the Recovery-based DG method \cite{vanLeer:2007tc}. For numerical stability, small hyperviscosity was added into the simulations, \eg as done in \Ref{Parker:2014fc}. Specifically, the terms $\smash{-2\nu (p_x^2+p_y^2) \xbar{W}+(\nu/2) \pd_y^2 \xbar{W} }$ and $\smash{-\nu \pd_y^4 U}$ with $\nu= 0.001$ were added to the right-hand side of \Eqs{eq:WKE_drift} and \eq{eq:WKE_zonal}, respectively \cite{foot:hyperviscosity}. }
\label{fig:plotU}
\end{figure*}

Let us assume that the characteristic wavelengths for ZFs and DWs are $\lambda_{\rm zf}$ and $\lambda_{\rm dw}$, respectively, and
\begin{gather}
 		\epsilon \doteq 
 			\mathrm{max}	\left(
 				\frac{\lambda_{\rm dw}}{\lambda_{\rm zf}},
	 			\frac{L_{\rm D}}{\lambda_{\rm zf}}
 				\right)  \ll 1.
  		\label{eq:GO_condition_1}
\end{gather}
Hence, the following estimates will be adopted:
\begin{equation}
	\begin{aligned}
		\pd_y \xbar{W} \backsim \lambda_{\rm zf}^{-1} \xbar{W},  & \quad
		\pd_{\vec{p}} \xbar{W} \backsim \lambda_{\rm dw} \xbar{W}, \\
		\pd_y H \backsim \lambda_{\rm zf}^{-1} H, & \quad
		\pd_{\vec{p}} H \backsim L_{\rm D} H,
	\end{aligned}
\end{equation}
where $H$ denotes both $\mc{H}$ and $\Gamma$. (The latter estimate is given for the \textit{maximum} of $\pd_{\vec{p}}H$, which is realized at $p \sim L_D^{-1}$  \cite{foot:barotropic}.) This gives
\begin{gather}
	\frac{\pd^n H}{\pd y^n}\,\frac{\pd^n \xbar{W}}{\pd p_y^n} 
		\backsim \left(\frac{\lambda_{\rm dw}}{\lambda_{\rm zf}}\right)^n \!\!H\xbar{W} \,			
		\lesssim \epsilon^n H\xbar{W},\\
	\frac{\pd^n H}{\pd p_y^n}\,\frac{\pd^n \xbar{W}}{\pd y^n}
		\backsim \left(\frac{L_{\rm D}}{\lambda_{\rm zf}}\right)^n \!\!H\xbar{W} \, 		
		\lesssim \epsilon^n H\xbar{W}.
\end{gather}
Then, using the lowest-order approximations of the Moyal products (\App{app:Weyl}), \Eqs{eq:phase_space} reduce to
\begin{subequations} \label{eq:WKE}
 	\begin{gather} 
 		\pd_t \xbar{W} 
     		= \{ \mc{H}, \xbar{W}  \} 
     		+ 2 \Gamma \xbar{W} 
     		+ \xbarlong{F} 
     		- 2 \mu_{\rm dw} \xbar{W}, 
     	\label{eq:WKE_drift} \\ 
 		\pd_t U+ \mu_{\rm zf} \, U 
     		= \frac{\pd}{\pd y}
       		\int \frac{\mathrm{d}^2 p}{(2\pi)^2}
       		\frac{p_x p_y \xbar{W}}{ p_{\rm D}^4}, 
     	\label{eq:WKE_zonal} 
 	\end{gather}
\end{subequations}
where $\{ \cdot, \cdot \}$ is the canonical Poisson bracket \eq{eq:Poisson_bracket} and
\begin{subequations} \label{eq:WKE_II}
	\begin{gather} 
 		\mc{H} \simeq -\beta p_x / p_{\rm D}^2 
 					+ p_x U 
 					+ p_x U'' / p_{\rm D}^2, 
 				\label{eq:WKE_Hamiltonian}  \\
 		\Gamma 	\simeq  \{ U'', p_x/p_{\rm D}^2\}/2 
 						= - p_x p_y U'''/ p_{\rm D}^4. 
 				\label{eq:WKE_Damping}
 	\end{gather}
\end{subequations}
One may recognize \Eq{eq:WKE_drift} as a variation of the WKE, so we attribute \Eqs{eq:WKE} and \eq{eq:WKE_II} as the \textit{WKE limit} of the Wigner-Moyal formulation. Clearly, $\mc{H}$ acts as the drifton ray Hamiltonian while $\Gamma$ acts as the corresponding dissipation rate. [The factors of two in \Eq{eq:WKE_drift} are due to the fact that $\xbar{W}$ is quadratic in the DW amplitude.] In other words, $\omega(y, \vec{p}, t) \doteq \mc{H} + i\Gamma-i\mu_{\rm dw}$ can be viewed as the local complex frequency of DWs with given wave vector~$\vec{p}$.

Notice that our WKE differs from the tWKE, which assumes a simpler dispersion of DWs; namely,
\begin{subequations} \label{eq:WKE_III}
	\begin{gather} 
	\mc{H} = - \beta p_x/p_{\rm D}^2 + p_x U, \\
	\Gamma = 0.
 	\end{gather}
\end{subequations}
Although the difference is only in the high-order derivatives of $U$, these terms remain important for various reasons. For example, in the Hamiltonian $\mc{H}$, $U''$ can be comparable to $\beta$ (as is sometimes the case in geophysics \cite{Vasavada:2005gs}). Also, consider the following. In isolated systems, the tWKE is $\pd_t \xbar{W} = \{\mc{H}, \xbar{W}\,\}$, so it conserves DW quanta, or, in other words, the DW enstrophy $\mc{Z}_{\rm dw}$ [\Eq{eq:Zdw}]. At the same time, the ZF enstrophy $\mc{Z}_{\rm zf}$ [\Eq{eq:Zzf}] generally evolves, so the total enstrophy $\mc{Z} = \mc{Z}_{\rm dw} + \mc{Z}_{\rm zf}$ does too. This is in contradiction with the gHME, which conserves $\mc{Z}$, and can lead to overestimating the ZF velocity and shear generated by DW turbulence \cite{foot:N}. In contrast to the tWKE, our formulation is free from such issues because \Eqs{eq:WKE} and \eq{eq:WKE_II} exactly conserve both $\mc{Z}$ and $\mc{E}$ (\App{app:ee4}). Note that, in order to retain this conservation property, it is necessary to keep both $U'''$ and $U''$ in \Eqs{eq:WKE_II}. In this sense, \textit{\Eqs{eq:WKE} and \eq{eq:WKE_II} represent the simplest GO model that is physically meaningful in the nonlinear regime}. This is in agreement with \Ref{foot:Parker}, where a similar conclusion was made based on comparing the linear zonostrophic instability rate predicted by the CE2. (As a note on terminology, \Ref{foot:Parker} refers to the tWKE [\Eqs{eq:WKE} and \eq{eq:WKE_III}] as the ``Asymptotic WKE," \ie the limit obtained when one assumes the ZFs are asymptotically large scale. Also, \Ref{foot:Parker} refers to \Eqs{eq:WKE} and \eq{eq:WKE_II} as ``CE2-GO.")

\begin{figure*}
\centering
\includegraphics[scale=0.7]{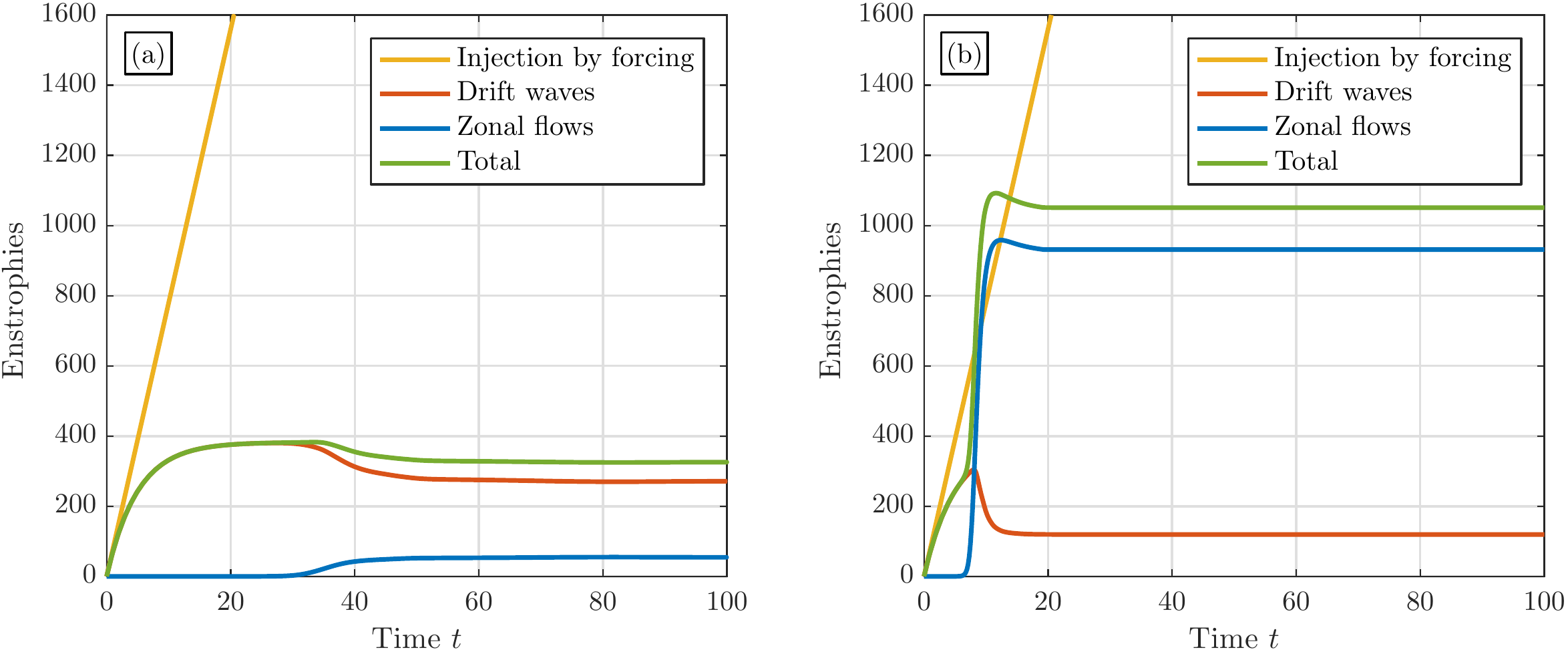}
\caption{The total, DW, and ZF enstrophies obtained by numerically integrating the WKE \eq{eq:WKE} for $\mc{H}$ and $\Gamma$ of two types: (a) our model [\Eqs{eq:WKE_II}]; (b) the tWKE model [\Eqs{eq:WKE_III}]. The yellow lines show the total enstrophy that one would get due to the external force $\xbarlong{F}$ at $\mu_{\rm dw, zf} = 0$. The initial conditions and simulation parameters are the same as in \Fig{fig:plotU}.}
\label{fig:enstrophy}
\end{figure*}

\begin{figure*}
\centering
\includegraphics[scale=0.7]{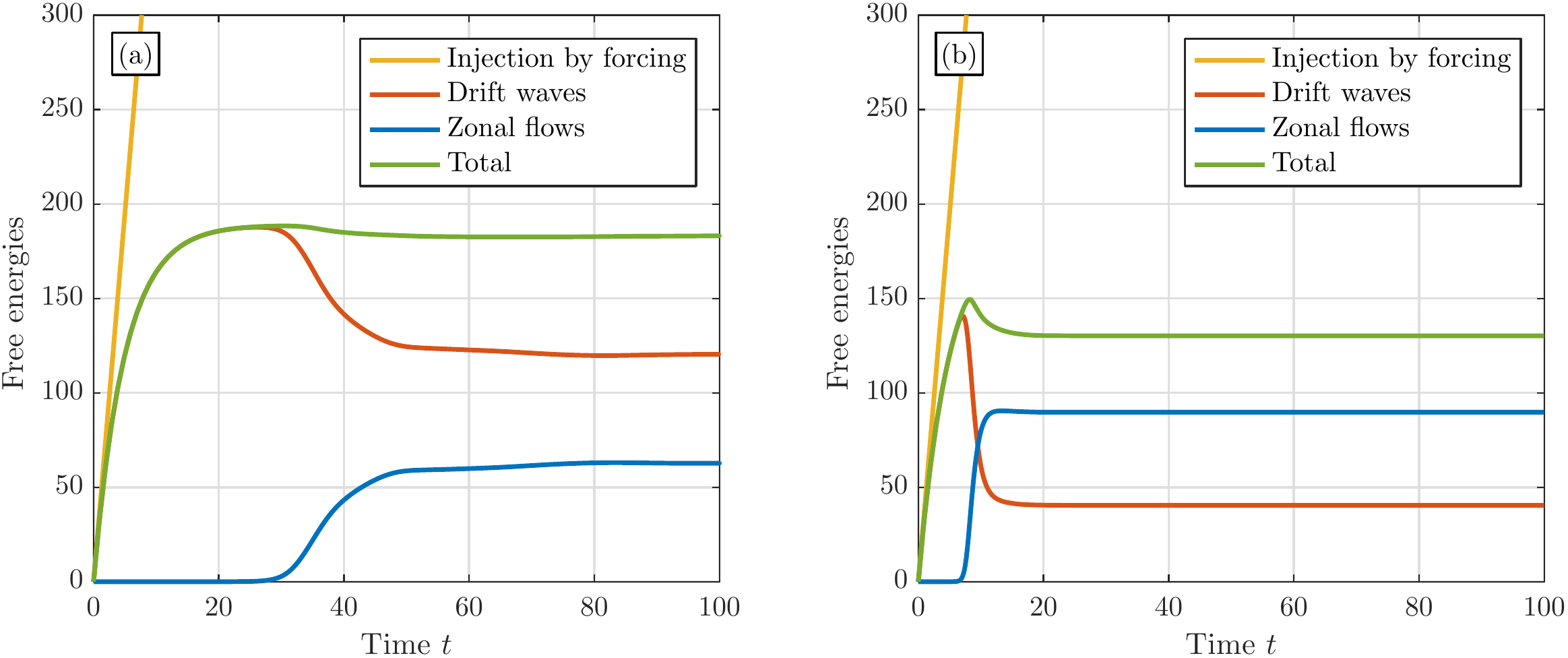}
\caption{The total, DW, and ZF energies obtained by numerically integrating the WKE \eq{eq:WKE} for $\mc{H}$ and $\Gamma$ of two types: (a) our model [\Eqs{eq:WKE_II}]; (b) the tWKE model [\Eqs{eq:WKE_III}]. The yellow lines show the total energy that one would get due to the external force $\xbarlong{F}$ at $\mu_{\rm dw, zf} = 0$. The initial conditions and simulation parameters are the same as in \Fig{fig:plotU}.}
\label{fig:energy}
\end{figure*}

The numerical results presented in Figs.~\ref{fig:plotU}-\ref{fig:energy} illustrate the importance of the difference between our WKE and the tWKE [subfigures (a) and (b), respectively]. As seen in \Fig{fig:plotU}, while our WKE model predicts ZFs with a particular $\lambda_{\rm zf}$, the scale of ZFs predicted by tWKE is determined by nothing but the grid size that is used in simulations. This is because the tWKE predicts that the rate of the zonostrophic instability $\gamma$ (\Sec{sec:growth}) scales linearly with the ZF wave number $q$, so ZFs are produced at the largest $q$ that is allowed (cf. \Ref{foot:Parker}).

Consider also the enstrophy plots in \Fig{fig:enstrophy}. To aid our discussion, we added plots of the enstrophy $\mc{Z}_{\rm ext}$ that the external forcing $\smash{\xbarlong{F}}$ injects into the DW-ZF system; namely, $\smash{\mc{Z}_{\rm ext} = (t/2) (2\pi)^{-2} \int \mathrm{d}y \, \mathrm{d}^2p \, \xbarlong{F}}$. Within our model, the total enstrophy $\mc{Z}$ remains always smaller than $\mc{Z}_{\rm ext}$, which is natural, since the simulation is done for $\mu_{\rm dw, zf} > 0$. In contrast, the tWKE model predicts that $\mc{Z}$ can surpass $\mc{Z}_{\rm ext}$, which is unphysical. In addition, the values of the ZF and total enstrophies predicted by the tWKE are several times larger than those predicted by our model. 

For the sake of completeness, \Fig{fig:energy} also presents the corresponding energies and the energy $\mc{E}_{\rm ext}$ introduced by the external force, $\smash{\mc{E}_{\rm ext} = (t/2)} \smash{(2\pi)^{-2} \int \mathrm{d}y \, \mathrm{d}^2p} \, \smash{\xbarlong{F} /p_{\rm D}^2}$. In both cases, $\mc{E}(t) \leq \mc{E}_{\rm ext}(t)$, which is in agreement with the fact that both models conserve the total energy of an isolated system. Still, the tWKE predicts very different results quantitatively even though the tWKE model [\Eqs{eq:WKE_III}] is seemingly close to ours [\Eqs{eq:WKE_II}].

\section{Conclusions}
\label{sec:conclusions}

The goal of this paper is to propose a new formulation of DW-ZF interactions that is more accurate than the tWKE and, simultaneously, more intuitive than the CE2. We adopt the same model [\Eqs{eq:QL3}] that was previously applied to derive the CE2. Then, we manipulate it using the Weyl calculus to produce a phase-space formulation of DW-ZF interactions. The resulting formulation [\Eqs{eq:phase_space} and \eq{eq:coefficients}] is akin to a quantum kinetic theory and involves a pseudodifferential Wigner--Moyal equation. To facilitate its numerical implementation in the future, we also present an integral representation of our main equations (\App{app:spectral}).

On one hand, this Wigner--Moyal formulation can be understood as an alternative representation to the CE2 since both models use the same assumptions. For example, we show that it leads to the same linear growth rate of weak ZFs as that obtained from the CE2 (\Sec{sec:growth}). On the other hand, the Wigner--Moyal formulation is arguably more intuitive than the CE2 for two reasons: (i) it permits treating driftons as particles (\ie as objects traveling in phase space), except now they are \textit{quantumlike} particles with nonzero wavelengths; and (ii) the separation between Hamiltonian effects and dissipation remains unambiguous even beyond the GO limit. 

Compared to the tWKE, the new approach is also more precise because (i)~it captures effects beyond the GO limit and (ii)~even in the GO limit, it predicts corrections to the tWKE that emerge from the newly found corrections to the drifton dispersion (\Sec{sec:WKE}). These corrections are essential as they allow DW-ZF enstrophy exchange, which is not included in the tWKE. By deriving the GO limit from first principles, we eliminate this discrepancy and arrive at a model that exactly conserves the total enstrophy (as opposed to the DW enstrophy conservation predicted by the tWKE) and the total energy, in agreement with the underlying gHME. We also illustrate the substantial difference between the GO limit of our WKE model and the tWKE using numerical simulations.

This work can be expanded at least in two directions. First, the difference between the Wigner--Moyal formulation and the newly proposed WKE can be assessed quantitatively using numerical simulations. Second, the analytic methods we proposed here can be extended to other turbulence models, such as those in \Refs{Anderson:2002ks, Anderson:2006cf}. The anticipated benefit is that more accurate equations would be derived that would respect fundamental conservation laws that existing theories may be missing otherwise.

The authors thank J.~A. Krommes for valuable discussions. This work was supported by the U.S. DOE through Contract Nos. DE-AC02-09CH11466 and DE-AC52-07NA27344, by the NNSA SSAA Program through DOE Research Grant No. DE-NA0002948, and by the U.S. DOD NDSEG Fellowship through Contract No. 32-CFR-168a.

\appendix

\section{Weyl calculus}
\label{app:Weyl}

This appendix summarizes our conventions for the Weyl transform. (For more information, see the excellent reviews in \Refs{Imre:1967fr,Tracy:2014to}.) The Weyl symbol $A(\vec{x}, \vec{p})$ of any given operator $\op{A}$ is defined as
\begin{gather}
 A(\vec{x}, \vec{p}) \doteq 
    \int \mathrm{d}^n s \, e^{-i \vec{p} \cdot \vec{s}} 
    \braket{ \vec{x}+ \vec{s}/2| \op{A} | \vec{x} - \vec{s}/2} .
\label{def:weyl_symbol}
\end{gather}
We shall refer to this description of the operators as a \textit{phase-space representation} since Weyl symbols are functions of the $2n$-dimensional ray phase space $(\vec{x}, \vec{p})$. Conversely, the inverse Weyl transformation is
\begin{multline}
 \op{A} = \frac{1}{(2\pi)^n} \int 
     \mathrm{d}^n x \, \mathrm{d}^n p \,  \mathrm{d}^n s \,
     e^{-i \vec{p}\cdot \vec{s}} \\
     \times A(\vec{x},\vec{p}) \ket{\vec{x}-\vec{s}/2} \bra{\vec{x}+\vec{s}/2}.
 \label{eq:weyl_inverse}
\end{multline}
In particular, notice that, for any operator $\op{A}$, its matrix elements in the coordinate representation, $\mathcal{A}(\vec{x}, \vec{x}') \doteq \braket{\vec{x}|\op{A}|\vec{x}'}$, can be expressed as
\begin{gather}
\mathcal{A}(\vec{x},\vec{x}') = \frac{1}{(2\pi)^n} \int
         \mathrm{d}^n p \,
        e^{-i \vec{p}\cdot (\vec{x}'-\vec{x})}
        A  \left(\frac{\vec{x}+\vec{x}'}{2}, \vec{p} \right),
 \notag
\end{gather}
so $A(\vec{x}, \vec{p})$ can be understood as a spectrum of $\mathcal{A}(\vec{x}, \vec{x}')$. In particular, 
\begin{gather} \label{eq:weyl_x_rep2}
\mathcal{A}(\vec{x},\vec{x}) = \int \frac{\mathrm{d}^n p}{(2\pi)^n}\, A  \left(\vec{x}, \vec{p} \right).
\end{gather}

Other properties of the Weyl transform that we use in this paper are as follows:

\begin{itemize}[leftmargin=*]
\item
As seen from \Eq{eq:weyl_x_rep2}, for any operator $\op{A}$, its trace $\mathrm{Tr}\,\op{A} \doteq \int \mathrm{d} x \, \braket{ x | \op{A} |x}$ can be expressed as
\begin{gather}
	\mathrm{Tr}\,\op{A} = \frac{1}{(2\pi)^n} \int \mathrm{d}^n x \, \mathrm{d}^n p\, A(\vec{x},\vec{p}).
 	\label{eq:trace}
\end{gather}
\item
If $A(\vec{x},\vec{p})$ is the Weyl symbol of $\op{A}$, then $A^*(\vec{x},\vec{p})$ is the Weyl symbol of $\op{A}^\dag$. As a corollary, the Weyl symbol of a Hermitian operator is real.
\item
For any $\op{C} = \op{A} \op{B}$, the corresponding Weyl symbols satisfy \cite{Moyal:1949gj,Groenewold:1946kp}
\begin{gather}
 C (\vec{x},\vec{p}) = A (\vec{x},\vec{p}) \star B (\vec{x},\vec{p}).
\label{eq:Moyal}
\end{gather}
Here, $\star$ is the \textit{Moyal product}, which is given by
\begin{gather}
  A (\vec{x},\vec{p}) \star B (\vec{x},\vec{p}) 
  \doteq 
  A (\vec{x},\vec{p}) e^{i \op{\mc{L}} /2} B (\vec{x},\vec{p}),
\label{def:Moyal}
\end{gather}
and $\op{\mc{L}}$ is the \textit{Janus operator}, which is given by
\begin{gather}
 \op{\mc{L}} \doteq 
   \overleftarrow{\pd_\vec{x}} \cdot \overrightarrow{\pd_\vec{p}}
   -  
   \overleftarrow{\pd_\vec{p}} \cdot \overrightarrow{\pd_\vec{x}}
    = \{ \cdot , \cdot \}.
\end{gather}
The arrows indicate the directions in which the derivatives act, and $A \hat{\mc{L}} B = \{A,B\}$ is the canonical Poisson bracket; namely,
\begin{gather}
 	\{ A , B \} \doteq 
  	(\pd_\vec{x} A) \cdot (\pd_\vec{p} B)
  	-
  	(\pd_\vec{p} A) \cdot (\pd_\vec{x} B).
 \label{eq:Poisson_bracket}
\end{gather} 
\item
The Moyal product is associative; \ie
\begin{gather}
	A \star B \star C \doteq (A \star B) \star C = A \star (B \star C).
\end{gather}
\item
The anti-symmetrized Moyal product defines the so-called \textit{Moyal bracket}
\begin{gather}
	\moysin{ A , B} 
    		\doteq - i\left(A \star B - B \star A \right) 
    		= 2 A \sin(\op{\mc{L}}/2)  B.
 \label{eq:sine_bracket}
\end{gather}
Because of the latter equality, this bracket is also called the \textit{sine bracket}. To lowest order,
\begin{gather}
	\moysin{ A , B} \simeq A\op{\mc{L}} B = \lbrace A , B \rbrace.
	\label{eq:ray_sin}
\end{gather}
\item
The symmetrized Moyal product is defined as
\begin{gather}
	 \moycos{ A , B} 
   		\doteq A \star B + B \star A 
 		= 2 A \cos(\op{\mc{L}}/2)  B.
 	\label{eq:cosine_bracket}
\end{gather}
Because of the latter equality, this bracket is also called the \textit{cosine bracket}. To lowest order,
\begin{gather}
	\moycos{A, B} \simeq 2AB.
	\label{eq:ray_cos}
\end{gather}
In a wave equation of the Wigner--Moyal type, such as \Eq{eq:Moyal_Liouville}, neglecting higher-order phase-space derivatives in the sine and cosine brackets leads to a WKE. Higher-order wave effects, such as diffraction and tunneling, are lost in this limit. For this reason, it is called the GO approximation, or the ray approximation.
\item
Assuming that fields vanish at infinity rapidly enough, the phase-space integral of the Moyal product of two symbols equals the integral of the regular product of these symbols; \ie
\begin{gather}
 	\int \mathrm{d}^n x \, \mathrm{d}^n p \, A \star B 
   		= \int \mathrm{d}^n x \, \mathrm{d}^n p \, A B . 
 	\label{eq:int_everywhere}
\end{gather}
As a corollary,
\begin{subequations} \label{eq:ints}
	\begin{gather}
 		\int \mathrm{d}^n x \, \mathrm{d}^n p \, \moysin{ A , B} 
 				= 0, \\
 		\int \mathrm{d}^n x \, \mathrm{d}^n p \, \moycos{ A , B} 
 				= 2 \int \mathrm{d}^n x \, \mathrm{d}^n p\, A B.
\end{gather}
\end{subequations}
\item For any constant $\vec{q}$, one has
\begin{align}
	A(\vec{p}) \star e^{i \vec{q}\cdot\vec{x}}
  	& = A(\vec{p}) e^{  \overleftarrow{\pd}_\vec{p} \cdot (\vec{q}/2)} 
  			e^{i \vec{q}\cdot\vec{x}}
  	\notag \\
  	& = A(\vec{p}+\vec{q}/2) e^{i \vec{q}\cdot\vec{x}}. \label{app:moyal_exp}
\end{align}

\item As a corollary, one has
\begin{subequations} \label{app:brackets_exp}
 	\begin{align}
  		\moysin{ A(\vec{p}&), e^{i \vec{q}\cdot\vec{x}}}
    		\notag \\
    		& = \frac{1}{i} \left[    A \left(\vec{p}+\frac{\vec{q}}{2} \right) 
           	-  A \left(\vec{p}-\frac{\vec{q}}{2} \right) \right]
     		e^{i \vec{q}\cdot\vec{x}}, \\
  		\moycos{ A(\vec{p}&), e^{i \vec{q}\cdot\vec{x}} }
    			\notag\\
    		& = \left[A \left(\vec{p}+\frac{\vec{q}}{2} \right) 
           		+  A \left(\vec{p}-\frac{\vec{q}}{2} \right) 
         		\right] e^{i \vec{q}\cdot\vec{x}} .
	\end{align}
\end{subequations}

\item For any constants $\vec{k}$ and $\vec{q}$, one can also show that
\begin{multline}
	A(\vec{p}) e^{i \vec{k}\cdot \vec{x}} \star B(\vec{p}) 
	e^{i\vec{q}\cdot \vec{x}} \\ 
	= A (\vec{p}+\vec{q}/2) B (\vec{p}-\vec{k}/2) 
	e^{i (\vec{k}+\vec{q}) \cdot \vec{x}}.
  	\label{app:moyal_exp2}
\end{multline}

\item Now we tabulate some Weyl transforms of various operators. (We use a two-sided arrow to show the correspondence with the Weyl transform.) First of all, the Weyl transforms of the identity, position, and momentum operators are given by
\begin{equation}
	\hat{1}	\,	\Leftrightarrow	\,	1, \quad
	\hat{x}^i \,	\Leftrightarrow	\, x^i, \quad
	\hat{p}_i \,	\Leftrightarrow	\, p_i.
\end{equation}
If $f$ and $g$ are any two functions, then
\begin{equation}
	f(\hat{\vec{x}}) 	\,	\Leftrightarrow	\, f(\vec{x}), \quad
	g(\hat{\vec{p}}) 	\,	\Leftrightarrow	\, g(\vec{p}).
\end{equation}
Similarly, using \Eq{def:Moyal}, we have
\begin{gather}
	f(\hat{\vec{x}}) \hat{p}_j		\,	\Leftrightarrow	\, p_j f(\vec{x})	+  (i/2) \pd_j 	f(\vec{x}), \\
	\hat{p}_j	f(\hat{\vec{x}}) 	\,	\Leftrightarrow	\, p_j f(\vec{x})	- (i/2) \pd_j 	f(\vec{x}).
\end{gather}
One may also notice the connection between these relations and the commutation relation between operators; \ie $[\hat{x}^j,  \hat{p}_k] = \hat{x}^j \hat{p}_k - \hat{p}_k \hat{x}^j  = i \delta^j_k$.  

\end{itemize}

\section{Spectral representation of the Wigner--Moyal formulation}
\label{app:spectral}

To facilitate numerical implementations of the Wigner--Moyal formulation in the future, we propose an integral form of \Eqs{eq:phase_space} using a spectral representation. (Numerical simulations of the CE2 theory were reported in \Refs{Farrell:2003dm, Farrell:2007fq, Marston:2008gx}.) The assumed notation for the Fourier representation of any $A(y, \vec{p}, t)$ will be
\begin{gather}\label{eq:WUq}
A(y,\vec{p},t) = \int \frac{\mathrm{d}q}{2\pi} \, A_q(\vec{p},t) e^{iqy}.
\end{gather}

We start by rewriting \Eqs{app:brackets_exp} as
\begin{align}
\mc{H} =  &   -\frac{\beta p_x}{p_{\rm D}^2} 
    + \int \frac{\mathrm{d}q}{2\pi} \, 
    \left(p_x e^{iqy} - \frac{q^2}{2}\,
    \moycos{ e^{iqy}, p_x p_{\rm D}^{-2} } \right)
    U_q \notag \\
 = &   -\frac{\beta p_x}{p_{\rm D}^2}  
    + \int \frac{\mathrm{d}q}{2\pi} \, 
      \left[ p_x - \frac{q^2}{2}
     \left(\frac{p_x}{p_{\mathrm{D},-q}^2} + \frac{p_x}{p_{\mathrm{D},+q}^2} \right)
     \right]
     U_q e^{iqy},\notag
\end{align}
where $p^2_{\mathrm{D}, \pm q} \doteq p_{\rm D}^2 (\vec{p} \pm \vec{e}_y q/2)$. This leads to
\begin{gather}
\mc{H}_q = 
    -2\pi  \delta(q) \,
     \frac{\beta p_x}{p_{\rm D}^2}
    + p_x \left[ 1 - \frac{q^2}{2}
     \left(\frac{1}{p_{\mathrm{D},-q}^2} + \frac{1}{p_{\mathrm{D},+q}^2} \right)
     \right] U_q.  \label{eq:H_q}
\end{gather}
Similarly,
\begin{multline}
 \Gamma = - \int \frac{\mathrm{d}q}{2\pi} \, 
       \frac{q^2}{2} 
      \,\moysin{ e^{iqy}, p_x p_{\rm D}^{-2}} \,U_q 
    \\
    =   -\int \frac{\mathrm{d}q}{2\pi} \, 
       \frac{q^2}{2i} 
      \left(\frac{p_x}{p_{\mathrm{D},-q}^2} - \frac{p_x}{p_{\mathrm{D},+q}^2} \right) U_q,  
\end{multline}
so one obtains
\begin{gather}
\Gamma_q = 
    \frac{i}{2} 
    \left(\frac{1}{p_{\mathrm{D},-q}^2} - \frac{1}{p_{\mathrm{D},+q}^2} \right)
    p_x q^2 U_q.  \label{eq:Gamma_q}
\end{gather}
Also, using \Eq{app:moyal_exp2}, we obtain
\begin{align}
 \moysin{ \mc{H}&,\xbar{W} \,}\notag\\
  &=  \int \frac{\mathrm{d}r \, \mathrm{d}s}{(2\pi)^2} \,
    \moysin{\mc{H}_r(\vec{p},t) e^{iry} , \xbar{W}_s (\vec{p},t) e^{isy}}  \notag\\
  &=  \int \frac{\mathrm{d}r \, \mathrm{d}s}{(2\pi)^2} \, 
    \frac{1}{i} 
    \left(
    \mc{H}_{r, +s} \xbar{W}_{s,-r} - \mc{H}_{r, -s} \xbar{W}_{s,+r}  
    \right) 
    e^{i(r+s)y},
 \notag
\end{align}
where $A_{r, \pm s} \doteq A_r (\vec{p} \pm \vec{e}_y s/2, t)$ for any $A_r(\vec{p},t)$. Also,
\begin{align}
\moycos{ \Gamma &,\xbar{W} \,}\notag\\
 & = \int \frac{\mathrm{d}r \, \mathrm{d}s}{(2\pi)^2} \,
     \moycos{\Gamma_r(\vec{p},t) e^{iry} , \xbar{W}_s (\vec{p},t) e^{isy}} \notag \\
 & = \int \frac{\mathrm{d}r \, \mathrm{d}s}{(2\pi)^2} \,
     \left(
     \Gamma_{r, +s} \xbar{W}_{s,-r} + \Gamma_{r, -s} \xbar{W}_{s,+r}  
     \right) 
     e^{i(r+s)y}. \notag
\end{align}
By inserting these expressions into \Eq{eq:Moyal_Liouville}, we obtain
\begin{multline}
\pd_t \xbar{W}_q = \xbarlong{F}_q - 2 \mu_{\rm dw} \xbar{W}_q \\
  +\int \frac{\mathrm{d}r}{2\pi} \, [
  \left(\Gamma_{r, +q-r} - i \mc{H}_{r, +q-r} \right) \xbar{W}_{q-r,-r} \\
  + \left(\Gamma_{r, +r-q} +i \mc{H}_{r, +r-q} \right) \xbar{W}_{q-r,r} 
  ]. \label{eq:phase_space_q}
\end{multline} 
Also, using that
\begin{align}
\frac{\pd}{\pd y}  &  \int \frac{\mathrm{d}^2 p}{(2\pi)^2}\,
        \frac{1}{ p_{\rm D}^2} \star p_x p_y \xbar{W} \star \frac{1}{ p_{\rm D}^2} \notag \\
       &  = \frac{\pd}{\pd y}  
        \int \frac{\mathrm{d}^2 p \, \mathrm{d}q}{(2\pi)^3}\,
        \frac{1}{ p_{\rm D}^2} \star 
        p_x p_y \xbar{W}_q e^{iqy} \star 
        \frac{1}{ p_{\rm D}^2} \notag \\
       &  = \frac{\pd}{\pd y}  
        \int \frac{\mathrm{d}^2 p \, \mathrm{d}q}{(2\pi)^3}\,
        \frac{p_x p_y}{ p_{\mathrm{D},+q}^2 p_{\mathrm{D},-q}^2} \, \xbar{W}_q e^{iqy} \notag \\
       &  = \frac{i}{2}
        \int \frac{\mathrm{d}^2 p \, \mathrm{d}q}{(2\pi)^3}
        \left(\frac{1}{p_{\mathrm{D},-q}^2} - \frac{1}{p_{\mathrm{D},+q}^2} \right)
        p_x \xbar{W}_q e^{iqy},
\end{align}
one gets the following representation of \Eq{eq:flow}:
\begin{gather}
\pd_t U_q
       + \mu_{\rm zf} U_q 
        = \frac{i}{2}
        \int \frac{\mathrm{d}^2 p \,}{(2\pi)^2} 
        \left(\frac{1}{p_{\mathrm{D},-q}^2} - \frac{1}{p_{\mathrm{D},+q}^2} \right)
        p_x \xbar{W}_q.
   \label{eq:flow_q}
\end{gather} 
Equations \eq{eq:H_q}, \eq{eq:Gamma_q}, \eq{eq:phase_space_q}, and \eq{eq:flow_q} constitute the spectral representation of our Wigner--Moyal formulation.

\begin{widetext}

\section{Conservation of the total enstrophy and energy}
\label{app:cons}

In this appendix we prove that in the case of isolated systems ($Q = 0$), the Wigner--Moyal and the WKE models conserve the total enstrophy $\mc{Z}$ and the total energy $\mc{E}$, whose expressions are given by \Eqs{eq:invariants} and \eq{eq:invariants_II}.

\subsection{Wigner--Moyal model}
\label{app:ee3}

First, consider the Wigner--Moyal model [\Eqs{eq:phase_space} and \eq{eq:coefficients}]. To show conservation of enstrophy, we obtain
\begin{align}
\frac{\mathrm{d}\mc{Z}}{\mathrm{d}t} 
 &  = \int \mathrm{d} y \, (\pd_y U) (\pd_y \pd_t U) 
    + 
    \frac{1}{2} \int
        \frac{\mathrm{d}^2 p}{(2\pi)^2} \,\mathrm{d}y \,
        \pd_t \xbar{W} \notag \\
 &  =  \frac{1}{2}
     \int \frac{\mathrm{d}^2p}{(2\pi)^2}\,\mathrm{d}y \,
     \left[
     2 U'''    
     \left(
      \frac{1}{ p_{\rm D}^2} \star p_x p_y \xbar{W} \star \frac{1}{ p_{\rm D}^2}
     \right)
     + \moysin{\mc{H}, \xbar{W}\,} + \moycos{ \Gamma , \xbar{W} \,}
     \right] \notag \\
 &  =  \frac{1}{2}
     \int \frac{\mathrm{d}^2p}{(2\pi)^2}\,\mathrm{d}y \,
     \left[2U'''    
     \left(
      \frac{1}{ p_{\rm D}^2} \star p_x p_y \xbar{W} \star \frac{1}{ p_{\rm D}^2}
     \right)
    + 2\Gamma \xbar{W} \,
    \right],
\end{align}
where we used \Eqs{eq:ints}. To evaluate the remaining terms, we use the Fourier representations of $\xbar{W}(y,\vec{p},t)$ and $U(y,t)$ as defined via \Eqs{eq:WUq}. Specifically, after substituting \Eq{eq:Damping} for $\Gamma$, we obtain
\begin{align}
\frac{\mathrm{d}\mc{Z}}{\mathrm{d}t}
   =  \frac{1}{2}
     \int 
     \frac{\mathrm{d}^2p}{(2\pi)^2}\,
     \frac{\mathrm{d}q}{2\pi} \, \frac{\mathrm{d}k}{2\pi}\,\mathrm{d}y \,
     U_q \xbar{W}_k
     \left[
      -2iq^3
       \left(
       \frac{1}{ p_{\rm D}^2} \star p_x p_y e^{iky} \star \frac{1}{ p_{\rm D}^2}
       \right) e^{iqy}
      - q^2 \moysin{ e^{iqy} , p_x p_{\rm D}^{-2}} e^{iky}
     \right].
\end{align}
The Moyal products and the brackets can be evaluated using \Eqs{app:moyal_exp} and \eq{app:brackets_exp}. Using the notation $A_{\pm q} \doteq A(\vec{p} \pm \vec{e}_y q/2)$ for any $A(\vec{p})$, one then obtains
\begin{align}
\frac{\mathrm{d} \mc{Z}}{\mathrm{d}t}
 &  =  \frac{1}{2i}
     \int \frac{ \mathrm{d}^2p}{(2\pi)^2}\,
     \frac{\mathrm{d}q}{2\pi} \, \frac{\mathrm{d}k}{2\pi} \, U_q \xbar{W}_k
     \left[ \frac{ 2 p_x p_y q^3}{p_{\mathrm{D},+k}^2 p_{\mathrm{D},-k}^2}
      - p_x q^2\left(
         \frac{1}{ p_{\mathrm{D},-q}^2} 
         -\frac{1}{ p_{\mathrm{D},+q}^2} 
        \right)  
     \right]
     \int \mathrm{d}y \, e^{i(k+q)y} \notag \\
 &  =  -i
     \int \frac{ \mathrm{d}^2p}{(2\pi)^2}\,
     \frac{\mathrm{d}q}{2\pi} \, \mathrm{d}k
     \, U_q \xbar{W}_k
       p_x p_y q^3 
       \left(
        \frac{1}{p_{\mathrm{D},+k}^2 p_{\mathrm{D},-k}^2} -
        \frac{1}{p_{\mathrm{D},+q}^2 p_{\mathrm{D},-q}^2}
       \right) \delta (k + q)	\notag \\
       & = 0.
\end{align}

To show conservation of energy, we obtain
\begin{align}
 \frac{\mathrm{d} \mc{E}}{\mathrm{d}t} 
  & =  \int \mathrm{d} y \, U (\pd_t U)
    + 
    \frac{1}{2} \int
        \frac{\mathrm{d}^2 p}{(2 \pi)^2} \,\mathrm{d} y\, 
        \frac{\pd_t \xbar{W}}{ p_{\rm D}^2} \notag \\
  & = \frac{1}{2}
     \int \frac{\mathrm{d}^2p}{(2\pi)^2} \,\mathrm{d}y\,
     \left[
     2U\,\frac{\pd}{\pd y}    
     \left(
      \frac{1}{ p_{\rm D}^2} \star p_x p_y \xbar{W} 
       \star \frac{1}{ p_{\rm D}^2} \right)
    + \frac{1}{ p_{\rm D}^2} \moysin{ \mc{H}, \xbar{W}\,} 
    + \frac{1}{ p_{\rm D}^2} \moycos{ \Gamma, \xbar{W}\,}  
    \right]\notag \\        
  & = -\frac{1}{2}
     \int \frac{\mathrm{d}^2p}{(2\pi)^2} \,\mathrm{d}y
     \left[ 2 U'  
     \left(
      \frac{1}{ p_{\rm D}^2} 
      \star p_x p_y \xbar{W} \star \frac{1}{ p_{\rm D}^2}
     \right)
    - \frac{1}{ p_{\rm D}^2} 
    \left\lbrace \!\left\lbrace p_x U 
    + \left[ \left[ U'' , p_x p_{\rm D}^{-2} \right] \right]/2
    , \xbar{W} \, 
    \right\rbrace \! \right\rbrace
    - \frac{1}{ p_{\rm D}^2} \moycos{ \moysin{ U'', p_x p_{\rm D}^{-2}}/2, \xbar{W} \,} 
    \right].
  \label{supp:aux}
\end{align}
Here, we used the fact that the Taylor expansion of \Eq{eq:sine_bracket} for the Moyal bracket $\smash{\moysin{ \xbar{W} , \beta p_x/p_{\rm D}^2}/p_{\rm D}^2}$ consists of total derivatives on~$y$, so its integral over $y$ is zero. The other terms can be expressed as follows. First of all,
\begin{align}
 \int \frac{\mathrm{d}^2p}{(2\pi)^2}\,\mathrm{d}y \, 
	  & \left[2 U'  
		   \left(
		     \frac{1}{ p_{\rm D}^2} \star p_x p_y \xbar{W} \star \frac{1}{ p_{\rm D}^2}
		   \right)
		   - \frac{1}{p_{\rm D}^2} \moysin{ p_x U, W} 
		   \right] \notag \\
	  & =\int \frac{\mathrm{d}^2p}{(2\pi)^2} \, 
		   \frac{\mathrm{d}q}{2\pi} \, \frac{\mathrm{d}k}{2\pi}\,\mathrm{d}y \, U_q 
		   \left[2 iq  \xbar{W}_k
		   \left(
		    \frac{1}{ p_{\rm D}^2} \star p_x p_y e^{iky} \star \frac{1}{ p_{\rm D}^2} 
		   \right) e^{iqy}
		   - \frac{1}{ p_{\rm D}^2} \moysin{ p_x e^{iqy}, \xbar{W}_k} e^{iky} \right] \notag \\
	  & =\int \frac{ \mathrm{d}^2p}{(2\pi)^2} \,  \frac{\mathrm{d}q}{2\pi} \, 
	  			\frac{\mathrm{d}k}{2\pi} \, U_q 
		   \left[ i  \xbar{W}_k
		    \frac{2 p_x p_y q}{ p_{\mathrm{D},+k}^2 p_{\mathrm{D},-k}^2}   
		   - \frac{p_x}{i p_{\rm D}^2} \left(\xbar{W}_{k,-q} - \xbar{W}_{k,+q} \right) \right] 
		   \int \mathrm{d}y \, e^{i(k+q)y} \notag \\
	  & =\int \frac{ \mathrm{d}^2p}{(2\pi)^2} \, \frac{\mathrm{d}q}{2\pi} \, 
	  			\mathrm{d}k \, i U_q \xbar{W}_k
		   \left[ 
		    \frac{2 p_x p_y q}{ p_{\mathrm{D},+k}^2 p_{\mathrm{D},-k}^2}   
		   + p_x \left(\frac{1}{ p_{\mathrm{D},+q}^2} 
		        - \frac{1}{p_{\mathrm{D},-q}^2}  \right) \right]
		   \delta(k + q) \notag \\
	 & = 0.\label{eq:aux101}
\end{align}
Also, using \Eq{app:moyal_exp2}, we obtain
\begin{align}
\int  & \frac{\mathrm{d}^2p}{(2\pi)^2} \,\mathrm{d}y\, 
			    \frac{1}{2 p_{\rm D}^2} \bigg(
			    \moysin{\moycos{U'', p_x p_{\rm D}^{-2}}, \xbar{W}\,}
			   +  
			    \moycos{\moysin{U'', p_x p_{\rm D}^{-2}}, \xbar{W}\,}
			  \bigg)
			   \notag \\
	  & = -
	   \int \frac{\mathrm{d}^2p}{(2\pi)^2} \, \frac{\mathrm{d}q}{2\pi} \, 
	   			\frac{\mathrm{d}k}{2\pi} \,\mathrm{d}y\,
			    \frac{U_q q^2}{2 p_{\rm D}^2} 
			    \bigg(
			    \moysin{\moycos{e^{iqy}, p_x p_{\rm D}^{-2}}, \xbar{W}_k e^{iky}}
			    +
			    \moycos{\moysin{e^{iqy}, p_x p_{\rm D}^{-2}}, \xbar{W}_k e^{iky}}
			    \bigg)
			    \notag \\
	  & = -
	   \int \frac{\mathrm{d}^2p}{(2\pi)^2} \, \frac{\mathrm{d}q}{2\pi} \, 
	   		\frac{\mathrm{d}k}{2\pi} \,\mathrm{d}y\,
		    \frac{ U_qp_x q^2}{2 p_{\rm D}^2} \bigg(
		    \moysin{(p_{\mathrm{D},+q}^{-2} + p_{\mathrm{D},-q}^{-2}) e^{iqy} , 
		    		\xbar{W}_k e^{iky}}
		    - \frac{1}{i}  \moycos{(p_{\mathrm{D},+q}^{-2} - p_{\mathrm{D},-q}^{-2}) e^{iqy} , 
		    			\xbar{W}_k e^{iky}}
		   \bigg)\notag\\
	  & =
	   \int \frac{\mathrm{d}^2p}{(2\pi)^2} \, \frac{\mathrm{d}q}{2\pi} \, 
	   			\frac{\mathrm{d}k}{2\pi} \,\mathrm{d}y\,
	    \frac{ U_qp_x q^2}{2i p_{\rm D}^2} \bigg(
	    \frac{\xbar{W}_{k,q}}{p_{\mathrm{D},+q-k}^2} 
	    	- \frac{\xbar{W}_{k,-q}}{p_{\mathrm{D},+q+k}^2}
	    + \frac{\xbar{W}_{k,q}}{p_{\mathrm{D},-q-k}^2} 
	    	- \frac{\xbar{W}_{k,-q}}{p_{\mathrm{D},-q+k}^2}
	    \notag \\ 
	  & \kern 150pt
	    + \frac{\xbar{W}_{k,q}}{p_{\mathrm{D},+q-k}^2} 
	    + \frac{\xbar{W}_{k,-q}}{p_{\mathrm{D},+q+k}^2}
	    - \frac{\xbar{W}_{k,q}}{p_{\mathrm{D},-q-k}^2} 
	    - \frac{\xbar{W}_{k,-q}}{p_{\mathrm{D},-q+k}^2}
	    \bigg)\,e^{i(k + q)y}\notag\\
	  & =
	   -i \int \frac{\mathrm{d}^2p}{(2\pi)^2} \, \frac{\mathrm{d}q}{2\pi} \, \mathrm{d}k\,
	    U_q \xbar{W}_{k} p_x q^2 \bigg(
	     \frac{1}{ p_{\mathrm{D},-q}^2 p_{\mathrm{D},-k}^2} 
	    - \frac{1}{ p_{\mathrm{D},+q}^2 p_{\mathrm{D},+k}^2}   
	 \bigg)\,\delta(k + q) \notag \\
	  & = 0.\label{eq:aux102}
\end{align}
By substituting \Eqs{eq:aux101} and \eq{eq:aux102} into \Eq{supp:aux}, one obtains $\dot{\mc{E}} = 0$.

\subsection{WKE model}
\label{app:ee4}

Now let us consider the WKE model [\Eqs{eq:WKE} and \eq{eq:WKE_II}]. To show conservation of enstrophy, we obtain
\begin{align}
\frac{\mathrm{d}\mc{Z}}{\mathrm{d}t}
	 &  = \int \mathrm{d} y \, (\pd_y U) (\pd_y \pd_t U) 
			    + \frac{1}{2} \int
			    \frac{\mathrm{d}^2 p}{(2\pi)^2}\,\mathrm{d} y \,
			    \pd_t \xbar{W} \notag \\
	 &  = \frac{1}{2} 
	    \int \frac{\mathrm{d}^2p}{(2\pi)^2} \,\mathrm{d} y 
			    \left(
			    \frac{2 p_x p_y}{ p_{\rm D}^4}\,U''' \xbar{W} 
			    + \{ \mc{H}, \xbar{W}\, \} + 2\Gamma \xbar{W} 
			    \right) \notag \\
	 &  = \frac{1}{2}
	    \int \frac{\mathrm{d}^2p}{(2\pi)^2} \,\mathrm{d} y 
			    \left(\frac{2 p_x p_y}{ p_{\rm D}^4} U''' + 2 \Gamma \right) \xbar{W}  
			    \notag \\
	 & = 0,
\end{align}
where we used \Eq{eq:WKE_Damping} and the fact that the integral of the Poisson bracket over all phase space is zero. In contrast, the tWKE does not conserve total enstrophy because $\Gamma=0$ [see \Eq{eq:WKE_III}]. This is also understood as follows: the tWKE manifestly conserves $\mathcal{Z}_{\rm dw}$, whereas $\mathcal{Z}_{\rm zf}$ is obviously not conserved; thus, $\mathcal{Z}_{\rm dw}+\mathcal{Z}_{\rm zf}$ cannot be conserved either. 

To show conservation of energy, we obtain
\begin{align}
 \frac{\mathrm{d}\mc{E}}{\mathrm{d}t} 
		  & =  \int \mathrm{d} y \, U (\pd_t U)
				    +  \frac{1}{2} \int \frac{\mathrm{d}^2 p}{(2 \pi)^2} \, \mathrm{d}y\,
				     \frac{\pd_t \xbar{W}}{p_{\rm D}^2} \notag \\
		  & = -\frac{1}{2}
				\int \frac{\mathrm{d}^2p}{(2\pi)^2} \,\mathrm{d}y
				     \frac{1}{p_{\rm D}^2}\left(
				    \frac{2 p_x p_y}{ p_{\rm D}^2} \, U' \xbar{W} 
				    - \{ \mc{H} , \xbar{W} \, \} 
				    - 2 \Gamma \xbar{W}
				    \right)  \notag \\
		  & = -\frac{1}{2}
		     \int \frac{\mathrm{d}^2p}{(2\pi)^2} \,\mathrm{d}y
				     \frac{1}{p_{\rm D}^2}\left(
				     \frac{2 p_x p_y}{ p_{\rm D}^2}\, U' \xbar{W}
				    - \left\{ 
				       p_x U + p_x p_{\rm D}^{-2}U'',
				       \xbar{W} \, \right\} 
				    + \frac{2 p_x p_y}{p_{\rm D}^4} U''' \xbar{W}\right),
\end{align}
where the integral of $\smash{\{ \xbar{W}, \beta p_x/ p_{\rm D}^2 \}/p_{\rm D}^2}$ over $y$ is zero because it can be written as a total derivative on~$y$. Finally,
\begin{align}
 \frac{\mathrm{d}\mc{E}}{\mathrm{d}t}
	  & = -\frac{1}{2}
			     \int \frac{\mathrm{d}^2p}{(2\pi)^2} \,\mathrm{d}y \, 
			    \left(
			           \frac{2 p_x p_y}{ p_{\rm D}^4} U' \xbar{W}
			    - \frac{p_x}{p_{\rm D}^2} U' \pd_{p_y} \xbar{W} 
			    -  \frac{ p_x}{p_{\rm D}^4}\, U'''\,\pd_{p_y} \xbar{W} 
			    -  \frac{ 2 p_x p_y}{p_{\rm D}^6}\, U'' \pd_y \xbar{W} 
			    + \frac{2 p_x p_y}{p_{\rm D}^6}\, U''' \xbar{W} 
			    \right)  \notag \\    
	  & = -\frac{1}{2}
	 \int \frac{\mathrm{d}^2p}{(2\pi)^2} \,\mathrm{d}y \, 
			    \left(
			      \frac{2 p_x p_y}{ p_{\rm D}^4}\, U' \xbar{W}
			    - \frac{2 p_x p_y}{p_{\rm D}^4} U' \xbar{W} 
			    -  \frac{ 4 p_x p_y}{p_{\rm D}^6}\, U''' \xbar{W} 
			    +  \frac{ 2 p_x p_y}{p_{\rm D}^6}\, U''' \xbar{W}
			    + \frac{2 p_x p_y}{p_{\rm D}^6}\, U''' \xbar{W}
			    \right)	\notag \\
	 &	= 0.
	 \label{eq:proof_energy}
\end{align}
Note that an analysis using the tWKE model leads to an expression similar to \Eq{eq:proof_energy} with only the first two terms in the integrand. These terms cancel out, thus showing that the tWKE model also conserves the total energy.

\end{widetext}
%



\begin{thebibliography}{44}
\expandafter\ifx\csname natexlab\endcsname\relax\def\natexlab#1{#1}\fi
\expandafter\ifx\csname bibnamefont\endcsname\relax
  \def\bibnamefont#1{#1}\fi
\expandafter\ifx\csname bibfnamefont\endcsname\relax
  \def\bibfnamefont#1{#1}\fi
\expandafter\ifx\csname citenamefont\endcsname\relax
  \def\citenamefont#1{#1}\fi
\expandafter\ifx\csname url\endcsname\relax
  \def\url#1{\texttt{#1}}\fi
\expandafter\ifx\csname urlprefix\endcsname\relax\def\urlprefix{URL }\fi
\providecommand{\bibinfo}[2]{#2}
\providecommand{\eprint}[2][]{\url{#2}}

\bibitem[{\citenamefont{G{\"u}rcan and Diamond}(2015)}]{Gurcan:2015jy}
\bibinfo{author}{\bibfnamefont{{\"O}.~D.} \bibnamefont{G{\"u}rcan}}
  \bibnamefont{and} \bibinfo{author}{\bibfnamefont{P.~H.}
  \bibnamefont{Diamond}}, ``{Zonal flows and pattern formation},''
  \bibinfo{journal}{J. Phys. A: Math. Theor.} \textbf{\bibinfo{volume}{48}},
  \bibinfo{pages}{293001} (\bibinfo{year}{2015}).

\bibitem[{\citenamefont{Vasavada and Showman}(2005)}]{Vasavada:2005gs}
\bibinfo{author}{\bibfnamefont{A.~R.} \bibnamefont{Vasavada}} \bibnamefont{and}
  \bibinfo{author}{\bibfnamefont{A.~P.} \bibnamefont{Showman}}, ``{Jovian
  atmospheric dynamics: an update after Galileo and Cassini},''
  \bibinfo{journal}{Rep. Prog. Phys.} \textbf{\bibinfo{volume}{68}},
  \bibinfo{pages}{1935} (\bibinfo{year}{2005}).

\bibitem[{\citenamefont{Johansen et~al.}(2009)\citenamefont{Johansen, Youdin,
  and Klahr}}]{Johansen:2009jf}
\bibinfo{author}{\bibfnamefont{A.}~\bibnamefont{Johansen}},
  \bibinfo{author}{\bibfnamefont{A.}~\bibnamefont{Youdin}}, \bibnamefont{and}
  \bibinfo{author}{\bibfnamefont{H.}~\bibnamefont{Klahr}}, ``{Zonal flows and
  long-lived axisymmetric pressure bumps in magnetorotational turbulence},''
  \bibinfo{journal}{Astrophys. J.} \textbf{\bibinfo{volume}{697}},
  \bibinfo{pages}{1269} (\bibinfo{year}{2009}).

\bibitem[{\citenamefont{Kunz and Lesur}(2013)}]{Kunz:2013jp}
\bibinfo{author}{\bibfnamefont{M.~W.} \bibnamefont{Kunz}} \bibnamefont{and}
  \bibinfo{author}{\bibfnamefont{G.}~\bibnamefont{Lesur}}, ``{Magnetic
  self-organization in Hall-dominated magnetorotational turbulence},''
  \bibinfo{journal}{Mon. Not. R. Astron. Soc.} \textbf{\bibinfo{volume}{434}},
  \bibinfo{pages}{2295} (\bibinfo{year}{2013}).

\bibitem[{\citenamefont{Diamond et~al.}(2005)\citenamefont{Diamond, Itoh, Itoh,
  and Hahm}}]{Diamond:2005br}
\bibinfo{author}{\bibfnamefont{P.~H.} \bibnamefont{Diamond}},
  \bibinfo{author}{\bibfnamefont{S.-I.} \bibnamefont{Itoh}},
  \bibinfo{author}{\bibfnamefont{K.}~\bibnamefont{Itoh}}, \bibnamefont{and}
  \bibinfo{author}{\bibfnamefont{T.~S.} \bibnamefont{Hahm}}, ``{Zonal flows in
  plasma{\textemdash}a review},'' \bibinfo{journal}{Plasma Phys. Control.
  Fusion} \textbf{\bibinfo{volume}{47}}, \bibinfo{pages}{R35}
  (\bibinfo{year}{2005}).

\bibitem[{\citenamefont{Fujisawa}(2009)}]{Fujisawa:2009jc}
\bibinfo{author}{\bibfnamefont{A.}~\bibnamefont{Fujisawa}}, ``{A review of
  zonal flow experiments},'' \textbf{\bibinfo{volume}{49}},
  \bibinfo{pages}{013001} (\bibinfo{year}{2009}).

\bibitem[{\citenamefont{Hillesheim et~al.}(2016)\citenamefont{Hillesheim,
  Delabie, Meyer, Maggi, Meneses, Poli, and {JET
  Contributors}}}]{EUROfusionConsortium:2016bk}
\bibinfo{author}{\bibfnamefont{J.~C.} \bibnamefont{Hillesheim}},
  \bibinfo{author}{\bibfnamefont{E.}~\bibnamefont{Delabie}},
  \bibinfo{author}{\bibfnamefont{H.}~\bibnamefont{Meyer}},
  \bibinfo{author}{\bibfnamefont{C.~F.} \bibnamefont{Maggi}},
  \bibinfo{author}{\bibfnamefont{L.}~\bibnamefont{Meneses}},
  \bibinfo{author}{\bibfnamefont{E.}~\bibnamefont{Poli}}, \bibnamefont{and}
  \bibinfo{author}{\bibnamefont{{JET Contributors}}}, ``{Stationary zonal flows
  during the formation of the edge transport barrier in the jet tokamak},''
  \bibinfo{journal}{Phys. Rev. Lett.} \textbf{\bibinfo{volume}{116}},
  \bibinfo{pages}{065002} (\bibinfo{year}{2016}).

\bibitem[{\citenamefont{Trines et~al.}(2005)\citenamefont{Trines, Bingham,
  Silva, Mendon{\c c}a, Shukla, and Mori}}]{Trines:2005in}
\bibinfo{author}{\bibfnamefont{R.}~\bibnamefont{Trines}},
  \bibinfo{author}{\bibfnamefont{R.}~\bibnamefont{Bingham}},
  \bibinfo{author}{\bibfnamefont{L.~O.} \bibnamefont{Silva}},
  \bibinfo{author}{\bibfnamefont{J.~T.} \bibnamefont{Mendon{\c c}a}},
  \bibinfo{author}{\bibfnamefont{P.~K.} \bibnamefont{Shukla}},
  \bibnamefont{and} \bibinfo{author}{\bibfnamefont{W.~B.} \bibnamefont{Mori}},
  ``{Quasiparticle approach to the modulational instability of drift waves
  coupling to zonal flows},'' \bibinfo{journal}{Phys. Rev. Lett.}
  \textbf{\bibinfo{volume}{94}}, \bibinfo{pages}{165002}
  (\bibinfo{year}{2005}).

\bibitem[{foo({\natexlab{a}})}]{foot:ray}
\bibinfo{note}{Specifically, this means that the traditional WKE has a unique
  set of characteristics at each phase-space location. Using the terminology
  that is adopted in the present paper, the ray approximation can be defined
  formally as a model based on the approximations \eq{eq:ray_sin} and
  \eq{eq:ray_cos}. This model is applicable under the assumption that ZFs have
  scales large compared to the characteristic wavelength of DWs, which is also
  known as the condition of geometrical optics. See below for details.}

\bibitem[{\citenamefont{Tracy et~al.}(2014)\citenamefont{Tracy, Brizard,
  Richardson, and Kaufman}}]{Tracy:2014to}
\bibinfo{author}{\bibfnamefont{E.~R.} \bibnamefont{Tracy}},
  \bibinfo{author}{\bibfnamefont{A.~J.} \bibnamefont{Brizard}},
  \bibinfo{author}{\bibfnamefont{A.~S.} \bibnamefont{Richardson}},
  \bibnamefont{and} \bibinfo{author}{\bibfnamefont{A.~N.}
  \bibnamefont{Kaufman}}, \emph{\bibinfo{title}{{Ray Tracing and Beyond: Phase
  Space Methods in Plasma Wave Theory}}} (\bibinfo{publisher}{Cambridge
  University Press}, \bibinfo{address}{New York}, \bibinfo{year}{2014}).

\bibitem[{foo({\natexlab{b}})}]{foot:Parker}
\bibinfo{note}{J. B. Parker, ``Dynamics of zonal flows: Failure of wave-kinetic
  theory, and new geometrical optics approximations," arXiv (2016),
  1604.06904v1, to appear in J. Plasma Phys.}

\bibitem[{\citenamefont{Farrell and Ioannou}(2003)}]{Farrell:2003dm}
\bibinfo{author}{\bibfnamefont{B.~F.} \bibnamefont{Farrell}} \bibnamefont{and}
  \bibinfo{author}{\bibfnamefont{P.~J.} \bibnamefont{Ioannou}}, ``{Structural
  stability of turbulent jets},'' \bibinfo{journal}{J. Atmos. Sci.}
  \textbf{\bibinfo{volume}{60}}, \bibinfo{pages}{2101} (\bibinfo{year}{2003}).

\bibitem[{\citenamefont{Farrell and Ioannou}(2007)}]{Farrell:2007fq}
\bibinfo{author}{\bibfnamefont{B.~F.} \bibnamefont{Farrell}} \bibnamefont{and}
  \bibinfo{author}{\bibfnamefont{P.~J.} \bibnamefont{Ioannou}}, ``{Structure
  and spacing of jets in barotropic turbulence},'' \bibinfo{journal}{J. Atmos.
  Sci.} \textbf{\bibinfo{volume}{64}}, \bibinfo{pages}{3652}
  (\bibinfo{year}{2007}).

\bibitem[{\citenamefont{Marston et~al.}(2008)\citenamefont{Marston, Conover,
  and Schneider}}]{Marston:2008gx}
\bibinfo{author}{\bibfnamefont{J.~B.} \bibnamefont{Marston}},
  \bibinfo{author}{\bibfnamefont{E.}~\bibnamefont{Conover}}, \bibnamefont{and}
  \bibinfo{author}{\bibfnamefont{T.}~\bibnamefont{Schneider}}, ``{Statistics of
  an unstable barotropic jet from a cumulant expansion},'' \bibinfo{journal}{J.
  Atmos. Sci.} \textbf{\bibinfo{volume}{65}}, \bibinfo{pages}{1955}
  (\bibinfo{year}{2008}).

\bibitem[{\citenamefont{Srinivasan and Young}(2012)}]{Srinivasan:2012im}
\bibinfo{author}{\bibfnamefont{K.}~\bibnamefont{Srinivasan}} \bibnamefont{and}
  \bibinfo{author}{\bibfnamefont{W.~R.} \bibnamefont{Young}}, ``{Zonostrophic
  instability},'' \bibinfo{journal}{J. Atmos. Sci.}
  \textbf{\bibinfo{volume}{69}}, \bibinfo{pages}{1633} (\bibinfo{year}{2012}).

\bibitem[{\citenamefont{Ait-Chaalal et~al.}(2016)\citenamefont{Ait-Chaalal,
  Schneider, Meyer, and Marston}}]{AitChaalal:2016jx}
\bibinfo{author}{\bibfnamefont{F.}~\bibnamefont{Ait-Chaalal}},
  \bibinfo{author}{\bibfnamefont{T.}~\bibnamefont{Schneider}},
  \bibinfo{author}{\bibfnamefont{B.}~\bibnamefont{Meyer}}, \bibnamefont{and}
  \bibinfo{author}{\bibfnamefont{J.~B.} \bibnamefont{Marston}}, ``{Cumulant
  expansions for atmospheric flows},'' \bibinfo{journal}{New J. Phys.}
  \textbf{\bibinfo{volume}{18}}, \bibinfo{pages}{025019}
  (\bibinfo{year}{2016}).

\bibitem[{\citenamefont{Farrell and Ioannou}(2009)}]{Farrell:2009ke}
\bibinfo{author}{\bibfnamefont{B.~F.} \bibnamefont{Farrell}} \bibnamefont{and}
  \bibinfo{author}{\bibfnamefont{P.~J.} \bibnamefont{Ioannou}}, ``{A stochastic
  structural stability theory model of the drift wave{\textendash}zonal flow
  system},'' \bibinfo{journal}{Phys. Plasmas} \textbf{\bibinfo{volume}{16}},
  \bibinfo{pages}{112903} (\bibinfo{year}{2009}).

\bibitem[{\citenamefont{Parker and Krommes}(2013)}]{Parker:2013hy}
\bibinfo{author}{\bibfnamefont{J.~B.} \bibnamefont{Parker}} \bibnamefont{and}
  \bibinfo{author}{\bibfnamefont{J.~A.} \bibnamefont{Krommes}}, ``{Zonal flow
  as pattern formation},'' \bibinfo{journal}{Phys. Plasmas}
  \textbf{\bibinfo{volume}{20}}, \bibinfo{pages}{100703}
  (\bibinfo{year}{2013}).

\bibitem[{\citenamefont{Parker and Krommes}(2014)}]{Parker:2014fc}
\bibinfo{author}{\bibfnamefont{J.~B.} \bibnamefont{Parker}} \bibnamefont{and}
  \bibinfo{author}{\bibfnamefont{J.~A.} \bibnamefont{Krommes}}, ``{Generation
  of zonal flows through symmetry breaking of statistical homogeneity},''
  \bibinfo{journal}{New J. Phys.} \textbf{\bibinfo{volume}{16}},
  \bibinfo{pages}{035006} (\bibinfo{year}{2014}).

\bibitem[{\citenamefont{Parker}(2014)}]{Parker:2014tb}
\bibinfo{author}{\bibfnamefont{J.~B.} \bibnamefont{Parker}}, Ph.D. thesis,
  \bibinfo{school}{Princeton University} (\bibinfo{year}{2014}).

\bibitem[{\citenamefont{Krommes and Kim}(2000)}]{Krommes:2000ec}
\bibinfo{author}{\bibfnamefont{J.~A.} \bibnamefont{Krommes}} \bibnamefont{and}
  \bibinfo{author}{\bibfnamefont{C.-B.} \bibnamefont{Kim}}, ``{Interactions of
  disparate scales in drift-wave turbulence},'' \bibinfo{journal}{Phys. Rev. E}
  \textbf{\bibinfo{volume}{62}}, \bibinfo{pages}{8508} (\bibinfo{year}{2000}).

\bibitem[{\citenamefont{Smolyakov and Diamond}(1999)}]{Smolyakov:1999jk}
\bibinfo{author}{\bibfnamefont{A.~I.} \bibnamefont{Smolyakov}}
  \bibnamefont{and} \bibinfo{author}{\bibfnamefont{P.~H.}
  \bibnamefont{Diamond}}, ``{Generalized action invariants for drift
  waves-zonal flow systems},'' \bibinfo{journal}{Phys. Plasmas}
  \textbf{\bibinfo{volume}{6}}, \bibinfo{pages}{4410} (\bibinfo{year}{1999}).

\bibitem[{\citenamefont{Moyal}(1949)}]{Moyal:1949gj}
\bibinfo{author}{\bibfnamefont{J.~E.} \bibnamefont{Moyal}}, ``{Quantum
  mechanics as a statistical theory},'' \bibinfo{journal}{Math. Proc. Cambridge
  Philos. Soc.} \textbf{\bibinfo{volume}{45}}, \bibinfo{pages}{99}
  (\bibinfo{year}{1949}).

\bibitem[{\citenamefont{Groenewold}(1946)}]{Groenewold:1946kp}
\bibinfo{author}{\bibfnamefont{H.~J.} \bibnamefont{Groenewold}}, ``{On the
  principles of elementary quantum mechanics},'' \bibinfo{journal}{Physica}
  \textbf{\bibinfo{volume}{12}}, \bibinfo{pages}{405} (\bibinfo{year}{1946}).

\bibitem[{foo({\natexlab{c}})}]{foot:Mendonca}
\bibinfo{note}{Related calculations were also proposed in
  \Refs{Mendonca:2011ix,Mendonca:2012fg}. However, the complete Wigner-Moyal
  equation for DW was not introduced explicitly in those papers, and the
  validity of its GO limit was not explored in detail. Here, we argue that such
  details are, in fact, crucial.}

\bibitem[{\citenamefont{Constantinou et~al.}(2014)\citenamefont{Constantinou,
  Farrell, and Ioannou}}]{Constantinou:2014fh}
\bibinfo{author}{\bibfnamefont{N.~C.} \bibnamefont{Constantinou}},
  \bibinfo{author}{\bibfnamefont{B.~F.} \bibnamefont{Farrell}},
  \bibnamefont{and} \bibinfo{author}{\bibfnamefont{P.~J.}
  \bibnamefont{Ioannou}}, ``{Emergence and equilibration of jets in beta-plane
  turbulence: applications of stochastic structural stability theory},''
  \bibinfo{journal}{J. Atmos. Sci.} \textbf{\bibinfo{volume}{71}},
  \bibinfo{pages}{1818} (\bibinfo{year}{2014}).

\bibitem[{foo({\natexlab{d}})}]{foot:qm}
\bibinfo{note}{The only peculiarity of our problem in this respect is that
  $\tilde{w}(\vec{x}, t)$ is real rather than complex. However, that is just a
  matter of initial conditions. The equation governing $\ket{\tilde{w}}$ still
  has a quantumlike form.}

\bibitem[{\citenamefont{Wigner}(1932)}]{Wigner:1932cz}
\bibinfo{author}{\bibfnamefont{E.}~\bibnamefont{Wigner}}, ``{On the quantum
  correction for thermodynamic equilibrium},'' \bibinfo{journal}{Phys. Rev.}
  \textbf{\bibinfo{volume}{40}}, \bibinfo{pages}{749} (\bibinfo{year}{1932}).

\bibitem[{foo({\natexlab{e}})}]{foot:quasi}
\bibinfo{note}{Beyond the ray approximation, the probability density of
  driftons cannot be defined, because wave quanta cannot be assigned specific
  locations in phase space due to the uncertainty principle. Unlike the true
  probability density, the Weyl symbol of the density matrix is not
  positive-definite in this case; however, it does remain real.}

\bibitem[{foo({\natexlab{f}})}]{foot:pseudo}
\bibinfo{note}{Equation \eq{eq:aux1} is a pseudo-differential equation as it
  contains, in general, phase-space derivatives of infinite order. For an
  extended discussion, see \Ref{McDonald:1988dp}.}

\bibitem[{\citenamefont{Lifshitz and Pitaevskii}(1981)}]{Lifshitz:1981ui}
\bibinfo{author}{\bibfnamefont{E.~M.} \bibnamefont{Lifshitz}} \bibnamefont{and}
  \bibinfo{author}{\bibfnamefont{L.~P.} \bibnamefont{Pitaevskii}},
  \emph{\bibinfo{title}{{Physical Kinetics}}} (\bibinfo{publisher}{Pergamon
  Press}, \bibinfo{address}{New York}, \bibinfo{year}{1981}).

\bibitem[{\citenamefont{Bakas et~al.}(2015)\citenamefont{Bakas, Constantinou,
  and Ioannou}}]{Bakas:2015iy}
\bibinfo{author}{\bibfnamefont{N.~A.} \bibnamefont{Bakas}},
  \bibinfo{author}{\bibfnamefont{N.~C.} \bibnamefont{Constantinou}},
  \bibnamefont{and} \bibinfo{author}{\bibfnamefont{P.~J.}
  \bibnamefont{Ioannou}}, ``{S3T Stability of the Homogeneous State of
  Barotropic Beta-Plane Turbulence},'' \bibinfo{journal}{J. Atmos. Sci.}
  \textbf{\bibinfo{volume}{72}}, \bibinfo{pages}{1689} (\bibinfo{year}{2015}).

\bibitem[{\citenamefont{Liu and Shu}(2000)}]{Liu:2000ee}
\bibinfo{author}{\bibfnamefont{J.-G.} \bibnamefont{Liu}} \bibnamefont{and}
  \bibinfo{author}{\bibfnamefont{C.-W.} \bibnamefont{Shu}}, ``{A high-order
  discontinuous-Galerkin method for 2D incompressible flows},''
  \bibinfo{journal}{J. Comput. Phys.} \textbf{\bibinfo{volume}{160}},
  \bibinfo{pages}{577} (\bibinfo{year}{2000}).

\bibitem[{\citenamefont{Gottlieb et~al.}(2001)\citenamefont{Gottlieb, Shu, and
  Tadmor}}]{Gottlieb:2001iy}
\bibinfo{author}{\bibfnamefont{S.}~\bibnamefont{Gottlieb}},
  \bibinfo{author}{\bibfnamefont{C.-W.} \bibnamefont{Shu}}, \bibnamefont{and}
  \bibinfo{author}{\bibfnamefont{E.}~\bibnamefont{Tadmor}}, ``{Strong
  stability-preserving high-order time discretization methods},''
  \bibinfo{journal}{SIAM Review} \textbf{\bibinfo{volume}{43}},
  \bibinfo{pages}{89} (\bibinfo{year}{2001}).

\bibitem[{\citenamefont{van Leer et~al.}(2007)\citenamefont{van Leer, Lo, and
  van Raalte}}]{vanLeer:2007tc}
\bibinfo{author}{\bibfnamefont{B.}~\bibnamefont{van Leer}},
  \bibinfo{author}{\bibfnamefont{M.}~\bibnamefont{Lo}}, \bibnamefont{and}
  \bibinfo{author}{\bibfnamefont{M.}~\bibnamefont{van Raalte}}, in
  \emph{\bibinfo{booktitle}{18th AIAA Computational Fluid Dynamics Conference}}
  (\bibinfo{address}{Miami}, \bibinfo{year}{2007}).

\bibitem[{foo({\natexlab{g}})}]{foot:hyperviscosity}
\bibinfo{note}{Decreasing the hyperviscosity parameter by an order of magnitude
  to $\nu = 0.0001$ leads to qualitatively similar results. Quantitatively, the
  steady-state ZF enstrophy $\mc{Z}_{\rm zf}$ increases by about $25\%$ when
  using the tWKE. This can be understood by how $U$ in the tWKE model is
  dominated by short wavelength components, which are more strongly affected by
  hyperviscosity.}

\bibitem[{foo({\natexlab{h}})}]{foot:barotropic}
\bibinfo{note}{In the barotropic limit $(L_{\rm D} \to \infty )$, the ordering
  \eq{eq:GO_condition_1} can no longer be satisfied. This means that a GO
  approximation of \Eqs{eq:phase_space} and \eq{eq:coefficients} is not
  possible in this limit. This problem originates from the fact that the
  Hamiltonian is divergent at $p=0$. In the unnatural case where $\xbar{W}=0$
  for $p < p_0$, then we can estimate $\partial_\vec{p} H \lesssim H / p_0$.
  Then, a GO approximation is possible if $( \lambda_{\rm zf} p_0)^{-1} \ll
  1$.}

\bibitem[{foo({\natexlab{i}})}]{foot:N}
\bibinfo{note}{The fundamental reason for this discrepancy is that the tWKE
  assumes $\xbar{W}$ to be the true distribution probability of driftons, but
  $\xbar{W}$ is not. The true probability distribution would satisfy a strictly
  conservative equation. In the GO limit, it can always be found, at least
  numerically, as will be discussed elsewhere. The special case when
  $\smash{\pd_t U''}$ is negligible is discussed in \Ref{foot:Parker}.}

\bibitem[{\citenamefont{Anderson et~al.}(2002)\citenamefont{Anderson, Nordman,
  Singh, and Weiland}}]{Anderson:2002ks}
\bibinfo{author}{\bibfnamefont{J.}~\bibnamefont{Anderson}},
  \bibinfo{author}{\bibfnamefont{H.}~\bibnamefont{Nordman}},
  \bibinfo{author}{\bibfnamefont{R.}~\bibnamefont{Singh}}, \bibnamefont{and}
  \bibinfo{author}{\bibfnamefont{J.}~\bibnamefont{Weiland}}, ``{Zonal flow
  generation in ion temperature gradient mode turbulence},''
  \bibinfo{journal}{Phys. Plasmas} \textbf{\bibinfo{volume}{9}},
  \bibinfo{pages}{4500} (\bibinfo{year}{2002}).

\bibitem[{\citenamefont{Anderson et~al.}(2006)\citenamefont{Anderson, Nordman,
  Singh, and Weiland}}]{Anderson:2006cf}
\bibinfo{author}{\bibfnamefont{J.}~\bibnamefont{Anderson}},
  \bibinfo{author}{\bibfnamefont{H.}~\bibnamefont{Nordman}},
  \bibinfo{author}{\bibfnamefont{R.}~\bibnamefont{Singh}}, \bibnamefont{and}
  \bibinfo{author}{\bibfnamefont{J.}~\bibnamefont{Weiland}}, ``{Zonal flow
  generation in collisionless trapped electron mode turbulence},''
  \bibinfo{journal}{Plasma Phys. Control. Fusion}
  \textbf{\bibinfo{volume}{48}}, \bibinfo{pages}{651} (\bibinfo{year}{2006}).

\bibitem[{\citenamefont{Imre et~al.}(1967)\citenamefont{Imre, {\"O}zizmir,
  Rosenbaum, and Zweifel}}]{Imre:1967fr}
\bibinfo{author}{\bibfnamefont{K.}~\bibnamefont{Imre}},
  \bibinfo{author}{\bibfnamefont{E.}~\bibnamefont{{\"O}zizmir}},
  \bibinfo{author}{\bibfnamefont{M.}~\bibnamefont{Rosenbaum}},
  \bibnamefont{and} \bibinfo{author}{\bibfnamefont{P.~F.}
  \bibnamefont{Zweifel}}, ``{Wigner method in quantum statistical mechanics},''
  \bibinfo{journal}{J. Math. Phys.} \textbf{\bibinfo{volume}{8}},
  \bibinfo{pages}{1097} (\bibinfo{year}{1967}).

\bibitem[{\citenamefont{Mendon{\c c}a and Hizanidis}(2011)}]{Mendonca:2011ix}
\bibinfo{author}{\bibfnamefont{J.~T.} \bibnamefont{Mendon{\c c}a}}
  \bibnamefont{and}
  \bibinfo{author}{\bibfnamefont{K.}~\bibnamefont{Hizanidis}}, ``{Improved
  model of quasi-particle turbulence (with applications to Alfv{\'e}n and drift
  wave turbulence)},'' \bibinfo{journal}{Phys. Plasmas}
  \textbf{\bibinfo{volume}{18}}, \bibinfo{pages}{112306}
  (\bibinfo{year}{2011}).

\bibitem[{\citenamefont{Mendon{\c c}a and Benkadda}(2012)}]{Mendonca:2012fg}
\bibinfo{author}{\bibfnamefont{J.~T.} \bibnamefont{Mendon{\c c}a}}
  \bibnamefont{and} \bibinfo{author}{\bibfnamefont{S.}~\bibnamefont{Benkadda}},
  ``{Nonlinear instability saturation due to quasi-particle trapping in a
  turbulent plasma},'' \bibinfo{journal}{Phys. Plasmas}
  \textbf{\bibinfo{volume}{19}}, \bibinfo{pages}{082316}
  (\bibinfo{year}{2012}).

\bibitem[{\citenamefont{McDonald}(1988)}]{McDonald:1988dp}
\bibinfo{author}{\bibfnamefont{S.~W.} \bibnamefont{McDonald}}, ``{Phase-space
  representations of wave equations with applications to the eikonal
  approximation for short-wavelength waves},'' \bibinfo{journal}{Phys. Rep.}
  \textbf{\bibinfo{volume}{158}}, \bibinfo{pages}{337} (\bibinfo{year}{1988}).

\end{thebibliography}
\end{document}